\documentclass[aps,showpacs,reprint,twocolumn,superscriptaddress]{revtex4-1}

\usepackage{grffile}
\usepackage{graphicx}
\usepackage{amsmath}
\usepackage{amssymb}
\usepackage{color}
\usepackage{verbatim}
\usepackage{float}
\usepackage{mathrsfs}
\usepackage{esint}
\usepackage{tabularx}
\usepackage{multirow}
\usepackage{braket}
\usepackage{color,soul}
\usepackage{hyperref}

\usepackage{array}
\newcolumntype{C}[1]{>{\centering\let\newline\\\arraybackslash\hspace{0pt}}m{#1}}

\begin{document}

\title{Impurity and soliton dynamics in a Fermi gas with nearest-neighbor interactions}
\author{A.-M. Visuri}
\affiliation{COMP Centre of Excellence, Department of Applied Physics, Aalto University, FI-00076 Aalto, Finland}
\affiliation{Department of Quantum Matter Physics, University of Geneva, 24 quai Ernest-Ansermet, 1211 Geneva, Switzerland}

\author{P. T\"{o}rm\"{a}}
\affiliation{COMP Centre of Excellence, Department of Applied Physics, Aalto University, FI-00076 Aalto, Finland}

\author{T. Giamarchi}
\email{Thierry.Giamarchi@unige.ch}
\affiliation{Department of Quantum Matter Physics, University of Geneva, 24 quai Ernest-Ansermet, 1211 Geneva, Switzerland}

\begin{abstract}

We study spinless fermions with repulsive nearest-neighbor interactions perturbed by an impurity particle or a local potential quench. Using the numerical time-evolving block decimation method and a simplified analytic model, we show that the pertubations create a soliton-antisoliton pair. If solitons are already present in the bath, the two excitations have a drastically different dynamics: The antisoliton
does not annihilate with the solitons and is therefore confined close to its origin while the soliton excitation propagates. We discuss the consequences for experiments with ultracold gases. 

\end{abstract}

\maketitle

\section{Introduction}

The properties of an impurity coupled to a bath are a paradigmatic problem of
many body physics. For classical baths, the diffusion of massive particles is described by Brownian motion. A quantum bath leads to more complex physics. In two and three dimensions, 
the problem is described in terms of quasiparticles in which the impurity is surrounded by
excitations of the bath. This is the case for polarons~\cite{Mahan_Many-particle_2000, Grusdt_New_2016}, where the bath is made of phonons, bosonic particles, or fermions such as in a Fermi liquid.
Novel effects exists in one dimension where the excitations of the bath are drastically affected by interactions. In particular, if the one-dimensional quantum bath
is a Tomonaga-Luttinger liquid~\cite{Giamarchi}, the massless excitations of the bath lead
to new diffusion properties for the impurity, such as a subdiffusion~\cite{Zvonarev_Spin_2007, Lamacraft_Dispersion_2009}, and behavior ranging from polaronic
to Anderson's orthogonality catastrophe~\cite{Kantian}.

Cold atomic gases provide an ideal testing ground for such nonequilibrium quantum many-body phenomena due to their weak coupling to the environment and tunable parameters. The motion of initially localized impurities of a different spin state~\cite{Kohl} and atom species~\cite{Catani} was measured in one-dimensional tubes in the continuum. Recent developments in experimental setups allow addressing and imaging atoms in optical lattices at the resolution of a single lattice site. This has made it possible to measure the dynamics of specific nonequilibrium many-body states~\cite{Hung_In_2014, Ott_Single_2016, Kuhr_Quantum-gas_2016}. One can for instance create a local energy shift by focusing a laser beam on a selected site and flip the spin of the atom at that site by a microwave pulse. These techniques were used for recording the time evolution of initially localized spin impurities~\cite{Fukuhara_Quantum_2013} and magnons~\cite{Fukuhara_Microscopic_2013} in bosonic rubidium.
The quantum gas microscope technique has also been extended to fermionic atoms~\cite{Cheuk_Quantum-Gas_2015, Parsons_Site-Resolved_2015, haller_single-atom_2015, Edge_Imaging_2015, Boll_Spin_2016, Drewes_Antiferromagnetic_2016}.

On the theoretical front, mobile impurities in homogeneous baths have been largely studied~\cite{Jaksch, Kamenev, Demler, Massel, Kantian, Grusdt, Sirker}. 
The situation is more complicated if the bath has a structure. In particular, long-range interactions can lead to a periodic arrangement of the bath particles.
New behaviors are possible in such systems, such as the localization of the impurity~\cite{Horovitz_Phase_2013}.
For a fermionic bath with nearest-neighbor interactions in the Mott insulator (MI) state, a diffusive motion of the impurity was predicted to be connected to soliton excitations in the bath~\cite{Visuri2}. This raises the question of how the impurity and the excitations behave when the ground state of the bath already contains solitons.
Soliton excitations occur in a variety of one- and quasi-one-dimensional compounds, such as quantum spin chains~\cite{Nagler_Propagating_1982, mourigal_fractional_2013, braun_emergence_2005, Umegaki_Spinon_2015}. Quantum gas experiments provide complementary systems and offer the possibility to measure spatially the motion of such excitations and the impurity.
These phenomena can thus be relevant in experiments with polar KRb molecules~\cite{Yan_Observation_2013, Hazzard_Many-body_2014}, Rydberg atoms~\cite{Barredo_Coherent_2015, labuhn_tunable_2016}, and magnetic atoms~\cite{dePaz_Nonequilibrium_2013, Baier_Extended_2016} in optical lattices, as well as ions in rf traps~\cite{Neyenhuis_Observation_2016}, where effects of long-range interactions have been observed.

In this article, we use a combination of numerical and analytic techniques to study the dynamics of an impurity coupled to a bath of fermions with repulsive nearest-neighbor interactions. We compare it to the simpler case of a local quench in potential. Although both perturbations create a soliton and an antisoliton excitation, we show that the antisoliton can form a bound state with the impurity, which drastically affects its motion. We consider baths in the MI state and at an incommensurate filling for which solitons are present in the ground state. In the latter case, energy conservation prevents the antisoliton excitation from annihilating with the solitons and further constrains its motion. We discuss the possibility of observing these phenomena in experiments with ultracold gases.

\section{The model}
\label{section:model}

\begin{figure}
\begin{center}
	\includegraphics[width=\linewidth]{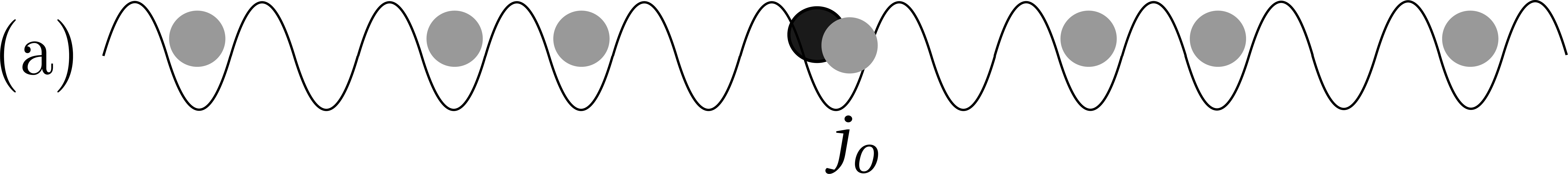}
	\includegraphics[width=\linewidth]{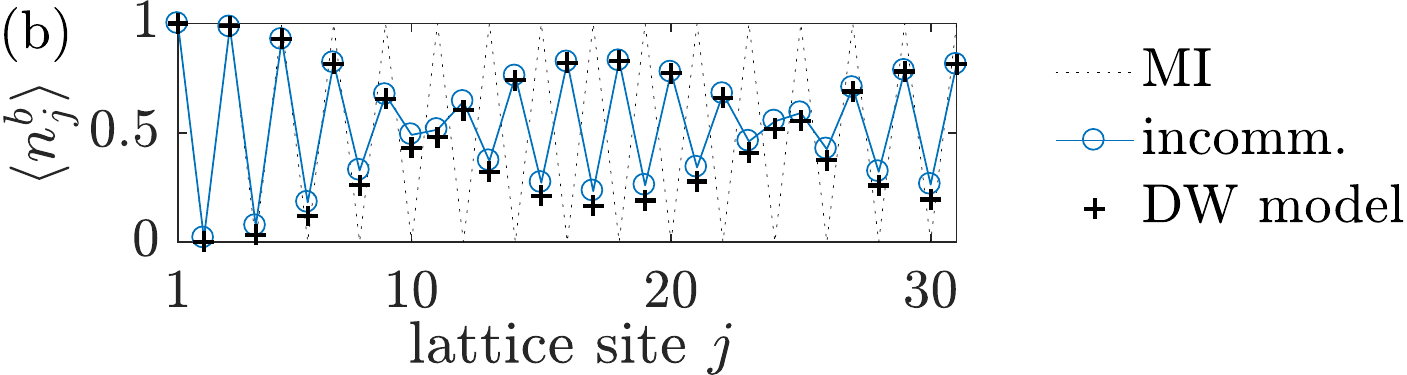}
	\caption{(a) Model: the bath contains spinless fermions (grey). At half filling, the ground state has a particle every two sites. Above half filling, solitons consisting of two neighboring 
        occupied sites exist, as schematically shown. At time $t = 0$, either an impurity particle (dark) or a static potential barrier is created at site $j_0$ at the center. (b) For incommensurate filling, the ground state density of the bath shows two neighboring maxima or minima at the most probable locations of the solitons. Only the left half of the lattice with $j \leq j_0$ is drawn, the other half is symmetric. The analytic result of our simplified domain wall (DW) model agrees well with the TEBD solution. The number of fermions is 31 for the MI and 33 for the incommensurate filling, $L = 61$, and $V = 50 J$ (see text).}
	\label{fig:ground_state}
\end{center}	
\end{figure}

We consider a bath described by 
\begin{equation}
H_{\text{b}} = -J \sum_{\langle i, j \rangle} b_{i}^{\dagger} b_{j} + V \sum_j \left(n_{j}^b - \frac{1}{2}\right)\left(n_{j + 1}^b - \frac{1}{2}\right),
\label{eq:bath_Hamiltonian}
\end{equation}
where $b_j$ ($b_j^\dagger$) annihilates (creates) a bath fermion at site $j$, $n_j^b=b_j^\dag b_j$ is the number operator of the bath fermions, $J$ is the tunneling amplitude, and $V$ is the nearest-neighbor interaction energy. We consider baths both in the MI state at commensurate (half) filling and at an incommensurate filling slightly above one half. The initial state is shown in the schematic drawing of Fig.~\ref{fig:ground_state}(a).
Our results also apply for the XXZ spin model, to which model (\ref{eq:bath_Hamiltonian}) can be mapped~\cite{Jordan_Uber_1928}. The dynamics of excitations in the XXZ model have been studied by the Bethe Ansatz~\cite{Vlijm_Spinon_2016}, time-evolving block decimation (TEBD)~\cite{Ganahl_Observation_2012, Visuri2}, and time-dependent DMRG~\cite{Langer_Real-time_2009, Langer_Real-time_2011} methods. The previous studies considered antiferromagnetic~\cite{Vlijm_Spinon_2016, Visuri2} or fully polarized~\cite{Ganahl_Observation_2012} initial states with one or a few flipped spins at the center. Ballistic and diffusive transport regimes were studied using different types of quenches~\cite{Langer_Real-time_2009, Langer_Real-time_2011}.

Here, we study the dynamics after either an impurity particle or a static potential barrier is introduced at the center site $j_0$ at time $t = 0$. 
The time evolution is described by the Hamiltonian $H~=~H_{\text{b}} + H^\prime$, where
\begin{align}
H^\prime &= -J \sum_{\langle i, j \rangle} c_{i}^{\dagger} c_{j} + U \sum_j \left(n_j^b - \frac{1}{2}\right) n_j, \label{eq:impurity_Hamiltonian}\\
H^\prime &= U n_{j_0}^b
\label{eq:static_potential_Hamiltonian}
\end{align}
respectively for the impurity and the potential. The repulsive on-site interaction between the impurity and the bath fermions is denoted by $U > 0$, the annihilation operator of the impurity by $c_j$, and $n_j = c_j^\dagger c_j$. The tunneling energies of the impurity and the bath fermions are equal. For the static potential barrier, the barrier height is equal to the on-site interaction energy.

\section{Results and discussion}
\label{sec:results}

We compute both ground-state and time-dependent observables with the numerical TEBD method~\cite{Daley_Time-dependent_2004, Vidal_Efficient_2004}. An odd number of lattice sites $L = 61$ and open boundary conditions make the ground state nondegenerate for commensurate filling with $N_b=31$ particles. For the incommensurate filling, $N_b=33$. A Schmidt number $\chi = 120$ is used in the imaginary time evolution. In the real time evolution, we use $\chi = 160$ and the second-order Trotter decomposition with time step $\delta t = 0.01/J$. The interaction energies are $U = V = 50 J$ unless mentioned otherwise. We focus on strong interactions to suppress the creation of solitons and antisolitons due to quantum fluctuations and to distinguish clearly the effects of the quenches. We show that essentially the same phenomena are observed also for interactions $U = V = 10 J$, which are closer to experimentally realizable values.

\subsection{Ground state density distribution}
\label{sec:ground_state}

In the MI state, for $V\gg J$, there is a particle every two sites and the density oscillates between $0$ and $1$. For incommensurate filling, the amplitude of the density oscillation decreases away from the boundaries. A low density of excess particles or holes with respect to half filling leads to soliton or antisoliton excitations~\cite{Giamarchi} where the phase of the density oscillation changes by $\pi$. The maxima of the soliton distribution due to the excess particles can be seen as domain walls (DWs) -- pairs of neighboring maxima or minima -- in the density distribution of Fig.~\ref{fig:ground_state}(b). For open boundary conditions, the solitons are not completely delocalized~\cite{weiss_finite_2008} as they would be for periodic boundary conditions. 

When the density of solitons is low and $V\gg J$, the solitons behave as free fermions~\cite{Giamarchi}. In this case, the density profile can be accurately predicted by a DW model, where one uses a bond representation to map the solitons to free fermions in an otherwise empty lattice (see Appendix~\ref{app:DW_model}). 
The wave function of $N$ free fermions can be written as
$\ket{\Psi_{1, 2, \cdots, N}} = \sum_{l_1, \cdots, l_N} \varphi_{k_1, k_2, \cdots, k_N}^{l_1, l_2, \cdots, l_N}
\times \ket{l_1} \otimes \ket{l_2} \otimes \cdots \ket{l_N}$,
where $l_{\alpha}$ is the coordinate of fermion $\alpha$. The single-particle state where site $l$ is occupied and the other sites are empty is denoted by $\ket{l} = \ket{0, \cdots, 0, 1_l, 0, \cdots, 0}$. The coefficient $\varphi_{k_1, \cdots, k_N}^{l_1, \cdots, l_N}$ is given by the Slater determinant formed of the single-particle wave functions $\varphi_{k_m}^{l} = \sqrt{\frac{2}{L}} \sin(k_m l)$,
where $k_m = \frac{m \pi}{L}$ and $m = 1, \cdots, N$. One can thus calculate the expectation value of the density $\langle n_j^b \rangle = \bra{\Psi_{1, \cdots, N}} n_j^b \ket{\Psi_{1, \cdots, N}}$ as
\begin{equation*}
\langle n_j^b \rangle = \frac{1}{2} \sum_{l_1, \cdots, l_N} |\varphi_{k_1, \cdots, k_N}^{l_1, \cdots, l_N}|^2 
\times \prod_{d = 1}^{j - 1} \left[ 2 \sum_{\alpha = 1}^N \delta_{d, l_{\alpha}} - 1 \right] + \frac{1}{2}.
\end{equation*}
More details are given in Appendix \ref{app:DW_model}. 
The result of the DW model shown in Fig.~\ref{fig:ground_state}(b) agrees very well with the numerical solution.

\subsection{Time evolution in the Mott insulator state}
\label{sec:commensurate_evolution}

\begin{figure}
\begin{center}
	\includegraphics[width=\linewidth]{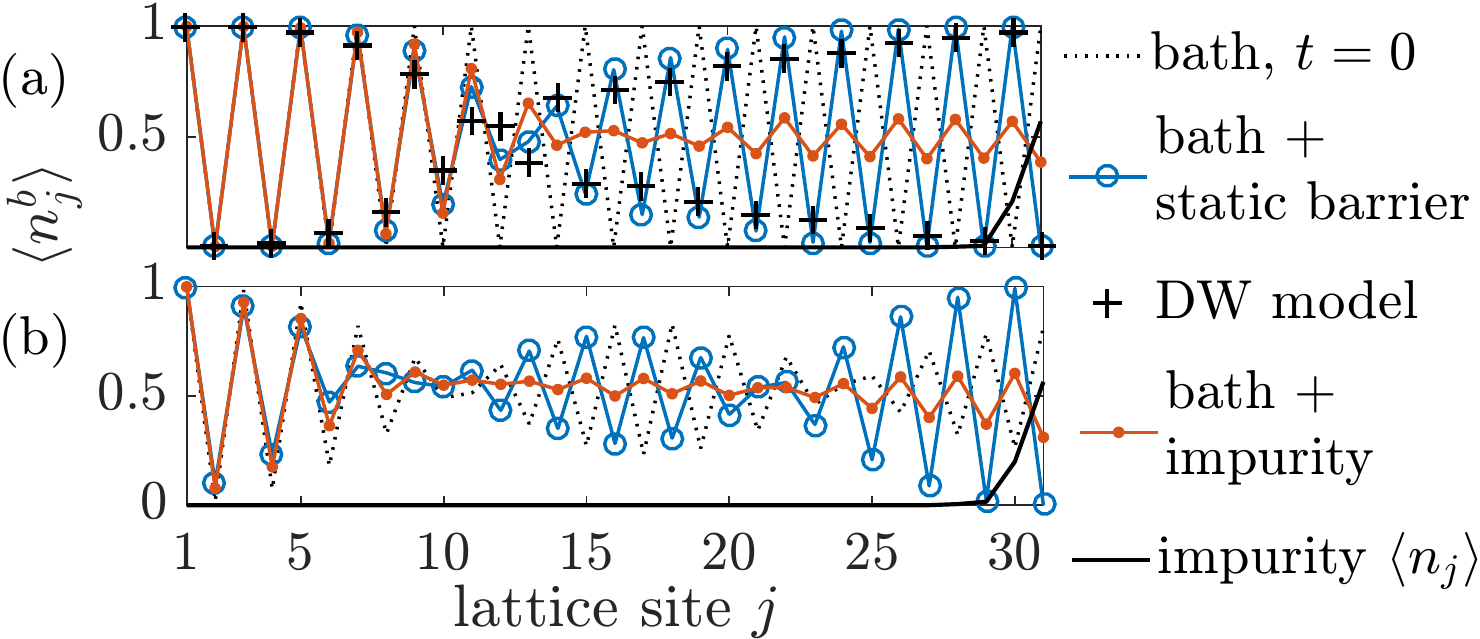}
	\caption{The density distribution of the bath $\langle n_j^b \rangle$ and the impurity $\langle n_j \rangle$ in the first half of the lattice. Panel (a) (resp. (b)) is for the commensurate (resp. incommensurate) filling. The bath density is shown at $t = 0$ and the bath and impurity densities at (a) $t = 6/J$, (b) $t = 8/J$. The analytic result of the DW model in panel (a) agrees well with the TEBD result.}
	\label{fig:symmetric_time_evolution}
\end{center}	
\end{figure}

We now turn to the time evolution. In the MI state, at $t=0$, the central site $j_0$ is occupied by a bath fermion with a high probability.  
A local potential quench [Eq. (\ref{eq:static_potential_Hamiltonian})] causes the fermion at $j_0$ to tunnel to the neighboring
site. A soliton and an antisoliton excitation are created in this process and propagate symmetrically in opposite directions. 
The density profiles in Fig.~\ref{fig:symmetric_time_evolution}
show an excitation as an additional DW. We can verify that the two excitations propagate symmetrically by studying the correlation of density on both sides of $j_0$, as is done in Appendix~\ref{app:density_correlation}. The symmetric propagation results in the complete inversion of the density profile between the excitations, as shown in Fig.~\ref{fig:symmetric_time_evolution}(a). The analytic result in Fig.~\ref{fig:symmetric_time_evolution}(a) is based on the DW model and agrees very well with the numerical solution. 

In the case of an impurity, the time evolution is richer. The impurity can oscillate between the two empty sites which form the antisoliton, and thus form a bound state with the antisoliton~\cite{Visuri2}. In this case, only the soliton excitation propagates and the density is inverted on one side of the lattice. The superposition of evolutions where both excitations propagate or only the soliton propagates in either direction results in a reduced amplitude of the density oscillation with respect to the initial state.
In addition, the impurity can move past a neighboring occupied site only in a second-order process with velocity $\frac{4J^2}{U}$ \cite{Georges_Strongly_2013}, which for $U \gg J$ results in the very slow motion of the impurity observed in Fig.~\ref{fig:symmetric_time_evolution}. For the incommensurate filling, the numerical results in Fig.~\ref{fig:symmetric_time_evolution}(b) show a similar population reversal in the regions between the DWs as in the MI. The DWs in the density remain in the final state after the excitation has passed. Using the DW model when solitons are present in the ground state is more delicate since the excitations can interact with these solitons (see Appendix~\ref{app:DW_model}) and a mapping to free fermions would be inaccurate. 

\begin{figure}
\begin{center}
	\includegraphics[width=\linewidth]{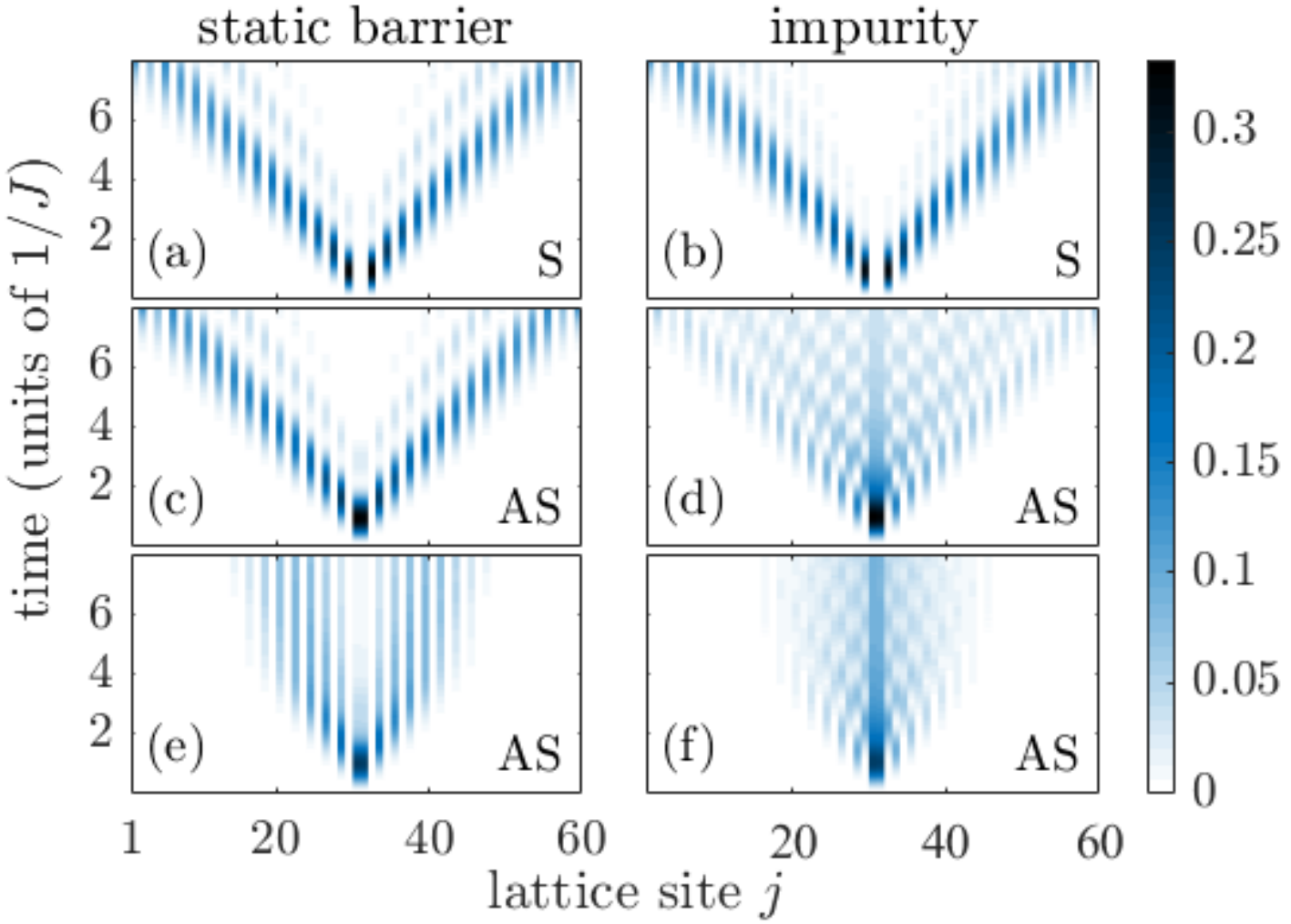}
	\caption{(a), (b): The soliton distribution $\langle n_j^b n_{j+1}^b \rangle$ as a function of position and time for the MI initial state when the system is perturbed with a static barrier or an impurity, respectively. The distributions are almost identical. (c), (d): The corresponding antisoliton distributions $\langle n_j^h n_{j+1}^h \rangle$. The antisoliton can form a bound state with the impurity (d), and a maximum remains at the center. (e), (f): The antisoliton distributions for incommensurate filling. The motion of the antisoliton is restricted because of the solitons present in the initial state.}
	\label{fig:commensurate_excitations}
\end{center}	
\end{figure}

In order to understand how the solitons in the initial state affect the new soliton and antisoliton excitations, we compute the distributions of neighboring filled sites $\langle n_j^b n_{j+1}^b \rangle$ and empty sites or holes $\langle n_j^h n_{j+1}^h \rangle = \langle (1 - n_j^b) (1 - n_{j+1}^b) \rangle$ as functions of time.
They correspond respectively to the distributions of solitons and antisolitons. Note that the measurement of such correlations is well within reach of experiments with quantum gas microscopes~\cite{Preiss_Strongly_2015}. 
The soliton distribution for the MI initial state is shown in panels (a) and (b) of Fig.~\ref{fig:commensurate_excitations}. The distribution is zero at $j_0-1$ and $j_0$, and obtains maxima at $j_0-2$ and $j_0+1$ when the bath fermion at $j_0$ tunnels either to $j_0-1$ or $j_0+1$. The distributions result from the superposition of these two configurations. Correspondingly, the antisoliton distributions in panels (c)--(f) have initially a maximum at $j_0-1$ and $j_0$.

\begin{figure}
\begin{center}
	\includegraphics[width=\linewidth]{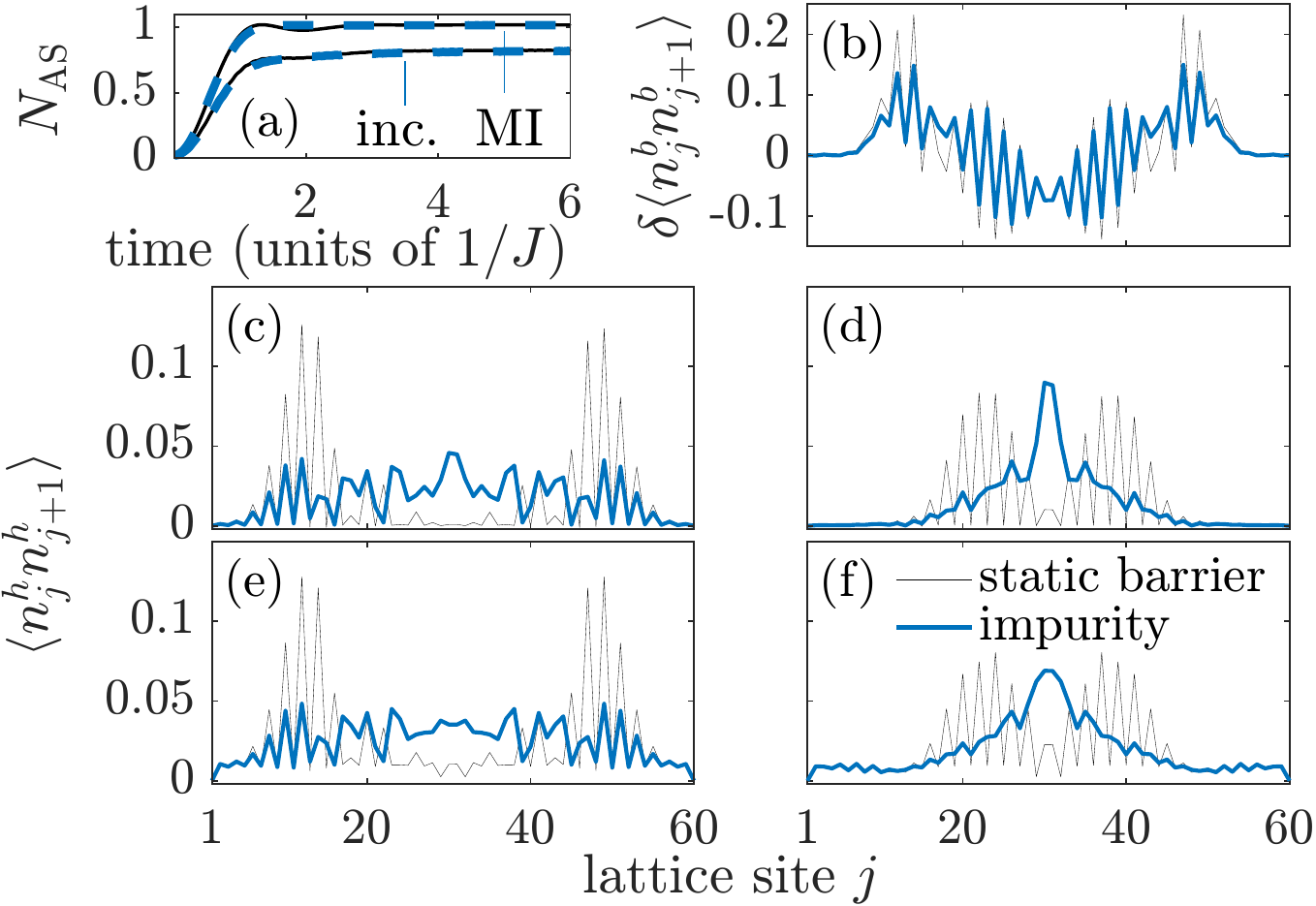}
	\caption{(a) The number of antisolitons $N_{\text{AS}}$ as a function of time for both perturbations and initial states. A dashed line is used for the impurity since the lines for the static barrier and impurity overlap. When the initial state is a MI (resp. incommensurate), $N_{\text{AS}}$ saturates at a value close to $1$ (resp. $0.8$). (b) The difference between the soliton distributions at times $t = 6/J$ and $t = 0$, $\delta \langle n_j^b n_{j+1}^b \rangle = \bra{\psi(t)} n_j^b n_{j+1}^b \ket{\psi(t)} - \bra{\psi(0)} n_j^b n_{j+1}^b \ket{\psi(0)}$, for incommensurate filling. The distributions are very close to each other for the two perturbations. (c)--(f): The antisoliton distribution $\langle n_j^h n_{j+1}^h \rangle$ at time $t~=~6/J$ for (c), (e) commensurate and (d), (f) incommensurate filling. For the impurity, the distribution has a maximum at the center whereas for the static barrier, there is a minimum. In panels (a)--(d), $U=V=50J$ and in panels (e) and (f), $U=V=10J$.}
	\label{fig:line_profiles}
\end{center}	
\end{figure}

The different perturbations lead to distinctly different dynamics. In Figs.~\ref{fig:commensurate_excitations}(a) and (c), where the potential is quenched at $j_0$ in the MI state, the distributions of both solitons and antisolitons become zero at the center, indicating that the two excitations indeed propagate symmetrically in opposite directions. The soliton distributions in Figs.~\ref{fig:commensurate_excitations}(a) and (b) are almost identical for the static barrier and the impurity. For the antisolitons, on the contrary, creating an impurity at $j_0$ produces an interference pattern and a maximum remains at the center, as seen in Fig.~\ref{fig:commensurate_excitations}(d). The interference pattern indicates that the antisoliton can be in several momentum states. Whereas in the potential quench, the soliton and antisoliton excitations have a high probability to obtain momenta of equal magnitude in opposite directions, in the case of the impurity, part of the momentum in the direction of the antisoliton can be absorbed by the impurity. The antisoliton can thus have a smaller momentum than the soliton. The impurity can form a bound state with the antisoliton, 
as shown by the remaining maximum at the center of the lattice.
These differences are also seen in the line profiles of Fig.~\ref{fig:line_profiles}(c) taken at time $t = 6/J$. The soliton distributions at $t = 6/J$ are very similar to the antisoliton distribution for the static barrier in Fig.~\ref{fig:line_profiles}(c).

\subsection{Time evolution at incommensurate filling}
\label{sec:incommensurate_evolution}

For incommensurate filling, an additional effect is observed. Figure~\ref{fig:line_profiles}(b) shows that the soliton excitation does not propagate as a free particle but is slowed down by the interactions with the other solitons in the system. Interestingly, the antisoliton excitation moves even less. After the initial propagation, it stays confined in the central region of the lattice, as shown in Figs.~\ref{fig:commensurate_excitations}(e) and (f) and Fig.~\ref{fig:line_profiles}(d). This confinement can be explained by energy conservation: the motion of an antisoliton excitation past a soliton would mean that they annihilate. The time evolution can be drawn schematically as
\begin{align*}
\ket{\psi(0)} = &\ket{\text{o} \; \; \; \text{x} \; \; \; \text{o} \; \; \; \text{o} \; \; \; \text{x} {}^{\curvearrowleft}\underline{\text{o}} \; \; \; \text{x} \; \; \; \text{o} \; \; \; \text{o} \; \; \; \text{x} \; \; \; \text{o}} \\
\rightarrow &\ket{\text{o} \; \; \; \text{x} {}^{\curvearrowleft}\text{o} \; \; \; \text{o} \; \; \; \text{o} \; \; \; \underline{\text{x}} \; \; \; \text{x} {}^{\curvearrowleft}\text{o} \; \; \; \text{o} \; \; \; \text{x} \; \; \; \text{o}} \\
\rightarrow &\ket{\text{o} \; \; \; \text{o} \; \; \; \text{x} \; \; \; \text{o} \; \; \; \text{o} \; \; \; \underline{\text{x}} \; \; \; \text{o} \; \; \; \text{x} \; \; \; \text{o}  \; \; \; \text{x} \; \; \; \text{o}},
\end{align*}
where the sites with (without) bath particles are denoted by o (x). The impurity is not drawn but is considered to stay at the underlined central site. When $U=V\gg J$ and kinetic energy is not taken into account, the total energy in the first configuration is $E=2V+U=3V$. On the second line, $E=3V$, whereas on the last line, where the antisoliton has annihilated on the right side, $E=2V$. The annihilation should have a very low probability for $V \gg J$ since the released energy cannot be absorbed as kinetic energy. Only a soliton can move past another soliton and conserve energy. Since the solitons in the initial state are not completely localized, the antisoliton distribution does not go to zero abruptly in Fig.~\ref{fig:commensurate_excitations} but rather diminishes smoothly. 

To verify that the antisoliton does not annihilate with the solitons, we compute the total number of solitons $N_{\text{S}}(t) = \sum_j \bra{\psi(t)} n_j^b n_{j+1}^b \ket{\psi(t)}$ and antisolitons
$N_{\text{AS}}(t) = \sum_j \bra{\psi(t)} n_j^h n_{j+1}^h \ket{\psi(t)}$. Figure~\ref{fig:line_profiles}(a) shows $N_{\text{AS}}(t)$, which is close to zero at $t = 0$ and increases until $t \approx 1/J$, the time scale for the tunneling of the bath particle away from $j_0$. The number saturates to a value close to $1$ in the MI state and to a smaller value at incommensurate filling. For $t \gtrsim 1/J$, we have checked that both $N_{\text{S}}(t)$ and $N_{\text{AS}}(t)$ stay constant, confirming that no annihilation takes place. The lower saturation value for incommensurate filling is due to the finite probability of site $j_0$ being initially empty, in which case excitations would not be created. Consistently, the value to which $N_{\text{AS}}(t)$ saturates is approximately the initial occupation probability $\langle n_{j_0}^b(0) \rangle$.

\section{Experimental realization with ultracold dipolar gases}

Lattice models with long-range interactions have so far been realized with bosonic dipolar atoms~\cite{dePaz_Nonequilibrium_2013, Baier_Extended_2016} and molecules~\cite{Yan_Observation_2013, Hazzard_Many-body_2014}. In one dimension, spinless fermions with nearest-neighbor interactions can be mapped to a spin model \cite{Jordan_Uber_1928}. Since spin systems can be mapped to hard-core bosons \cite{Giamarchi}, bosonic particles with both hard-core and long-range interactions could be used to realize the fermionic bath studied here. 

In quantum gas experiments, $\frac{V}{J}$ between zero and approximately $2$ was measured in extended Bose-Hubbard~\cite{Baier_Extended_2016} and $t-J$-like~\cite{dePaz_Nonequilibrium_2013} models realized with magnetic atoms. In optical lattices,  $U$ is tunable by Feshbach resonances~\cite{inouye_observation_1998} and $J$ by the lattice spacing and depth, which allows to tune $\frac{V}{J}$~\cite{Cugliandolo}. 
For the same atoms and laser wavelengths as in~\cite{Baier_Extended_2016}, we estimate that a larger lattice depth would allow to reach $V \approx 10 J$, where $J \approx 2.7$ Hz~\cite{Visuri2}. A coherent Bose-Einstein condensate was preserved for up to 1 s \cite{Baier_Extended_2016}, which gives a time scale $t \approx 2.7 \frac{1}{J}$ sufficiently long to observe the different dynamics resulting from the two different perturbations. The time scale required to observe the confinement of the antisoliton at an incommensurate filling of the bath is of the same order of magnitude. Another possibility to realize the type of bath studied here are polar molecules confined to deep lattices. Spin exchange by dipole-dipole interaction was demonstrated with immobile KRb molecules using different rotational states as pseudospin states~\cite{Yan_Observation_2013, Hazzard_Many-body_2014}.

The results of Sec.~\ref{sec:results} have been obtained for a large repulsion for clarity. 
Figures~\ref{fig:line_profiles}(e) and (f) show that the same clear differences are observable for $U=V=10J$, although the probability of finding neighboring empty sites in the ground state is larger. The larger effective tunneling energy of the impurity leads to a broader maximum of the antisoliton distribution at the center of the lattice. Figures corresponding to Figs.~\ref{fig:symmetric_time_evolution}--\ref{fig:line_profiles}(a) are presented in Appendix~\ref{app:smaller_interaction} for $U=V=10J$. 

Excitations to higher bands can be prevented by making the band gap an order of magnitude larger than the tunneling energies and interactions~\cite{dePaz_Nonequilibrium_2013, Baier_Extended_2016}. We therefore expect choosing interactions $U= V\approx 10J$ larger than the bandwidth of the lowest band and smaller than the band gap to be feasible so that the single-band approximation is valid.
We consider a system initially in the ground state at zero temperature, whereas in experiments, the temperature of the gas cloud is nonzero. For temperatures close to or larger than $V$, thermally excited soliton-antisoliton pairs could have an effect on the dynamics. To minimize the effects of thermal excitations, post-selection techniques~\cite{Fukuhara_Quantum_2013} could be used. Note that initial-state preparation can be used to create an excited initial state with alternating occupation \cite{Schreiber_observation_2015}.

In experiments with ultracold gases, uniform box potentials have recently been realized~\cite{Meyrath_Bose-Einstein_2005, van_Es_Box_2010, Gaunt_Bose-Einstein_2013, Mukherjee_Homogeneous_2016}. Most experiments however use a harmonic potential for confining the gas cloud. We show in Appendix~\ref{app:trap} that a Mott insulator ground state can be realized in the central region of a sufficiently shallow harmonic trap. The dynamics in this case is the same as in a box trap. To realize an incommensurate phase, a box trap is required. On the other hand, superimposing a harmonic potential and a box leads to a stronger confinement of the antisoliton than a uniform box potential, as shown in Appendix~\ref{app:trap}.

\section{Conclusions}
\label{sec:conclutions}

In summary, we observe that the presence of solitons in the bath leads to very different dynamics of the soliton and antisoliton excitations created by a local perturbation. The antisoliton does not annihilate with the solitons and is therefore confined close to its origin while the soliton excitation propagates. This is an example of the restrictions imposed by energy conservation on the dynamics: For interactions larger than the bandwidth, the energy released in an annihilation could not be absorbed as kinetic energy. Besides the numerical results, the simplified analytic model developed here offers a basis for understanding the dynamics of soliton excitations which occur in various physical systems.

\begin{acknowledgments}
This work was supported by the Academy of Finland through its Centres of Excellence Programme (2012-2017) and under Project Nos. 263347, 251748, and 272490, and by the European Research  Council (ERC-2013-AdG-340748-CODE). Computing resources were provided by CSC--the Finnish IT Centre for Science and the Aalto Science-IT Project. This work was supported in part by the Swiss NSF under Division II and by the ARO-MURI Non-equilibrium Many-body Dynamics grant (W911NF-14-1-0003).
\end{acknowledgments}

\appendix

\section{Simplified domain wall model}
\label{app:DW_model}

Using a dual bond representation, one can build a simplified model of the soliton excitations. Here, we consider the ground state for the commensurate and incommensurate filling with a low density of excess particles. In this case, the solitons can be assumed to not interact. A system with filling slightly below one half, where antisolitons exist in the ground state, could be treated in the same way. We only consider dynamics in the case of a local potential quench and the MI initial state, where the soliton or antisoliton excitation can be described in terms of a single particle. Extending the model to account for the motion of the impurity or dynamics in the case of solitons in the ground state would require taking the interactions into account.

\subsection{Dual bond representation}
\label{sec:duality}

\begin{figure}
\begin{center}
	\includegraphics[width=\linewidth]{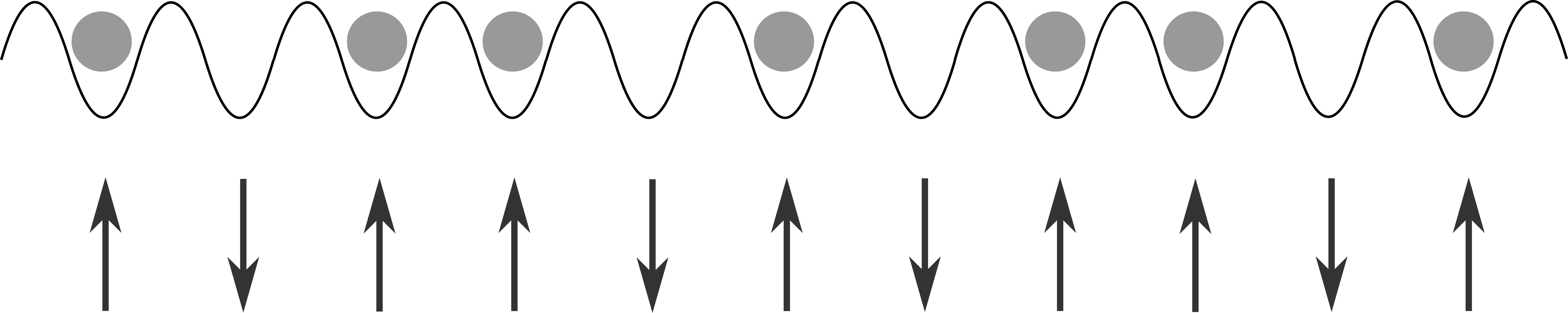}
	\caption{Spinless fermions with nearest-neighbor interactions off half filling can be mapped to an XXZ chain with a nonzero magnetization. The solitons of two neighboring occupied sites are mapped to two neighboring up spins.}
	\label{fig:ground_state_schematic}
\end{center}	
\end{figure}

The model of spinless fermions with nearest-neighbor interactions [Eq. (\ref{eq:bath_Hamiltonian}) of the main text] can be mapped to the XXZ spin model
\begin{equation}
H_{\text{b}} = J_{xy} \left( \frac{1}{2}\sum_{\langle i, j \rangle} S_i^+ S_j^- + \Delta \sum_j S_j^z S_{j + 1}^z \right)
\label{eq:XXZ_model}
\end{equation}
by the Jordan-Wigner transformation \cite{Jordan_Uber_1928}.
The mapping is illustrated in Fig. \ref{fig:ground_state_schematic}.
In the first sum of Eq. (\ref{eq:XXZ_model}), $\langle i, j \rangle$ denotes neighboring sites. The coupling constant between the spins is denoted by $J_{xy}$ and the anisotropy in the $z$ direction by $\Delta$. The operators $S_j^{\pm}~=~S_j^x~\pm~i S_j^y$ are the raising and lowering operators on site $j$ of a spin-$\frac{1}{2}$ chain, $S_j^{a} = \frac{1}{2} \sigma_j^{a}$, and $\sigma_j^{a}$ are the Pauli matrices with $a = x, y, z$. The parameters in terms of $J$ and $V$ are
\begin{equation}
\begin{split}
 J_{xy} &= 2 J, \\
 \Delta &= \frac{V}{J_{xy}}.
\end{split}
\end{equation}

\begin{figure}
\begin{center}
	\includegraphics[width=\linewidth]{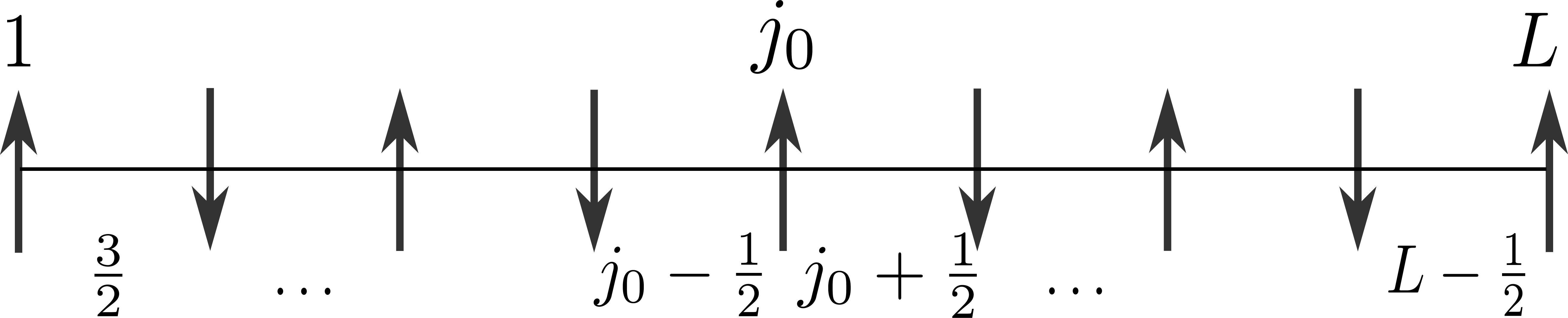}
	\caption{The indexing of the lattice sites and bonds of the spin model.}
	\label{fig:bond_indices}
\end{center}	
\end{figure}

The soliton excitations correspond in the XXZ model to two neighboring spins up, and antisolitons to two neighboring spins down. These excitations can represented by filled bonds, which are then mapped to spinless fermions in an otherwise empty lattice. One starts by replacing the lattice site operators in Hamiltonian (\ref{eq:XXZ_model}) by bond operators. In terms of the Pauli matrices $\sigma^a$, $H_b$ is written as
\begin{equation}
H_b = \frac{J_{xy}}{4} \sum_{j = 1}^{L - 1} \left( \sigma_j^x \sigma_{j + 1}^x + \sigma_j^y \sigma_{j + 1}^y + \Delta \sigma_j^z \sigma_{j + 1}^z \right).
\label{eq:H0_Pauli}
\end{equation}
In the spin notation, density operator is written as 
\begin{equation*}
n_j^b = S_j^z + \frac{1}{2} = \frac{1}{2} \sigma_j^z + \frac{1}{2}.
\end{equation*}
The local potential term of Eq. (\ref{eq:static_potential_Hamiltonian}) of the main text therefore transforms into
\begin{equation}
H^\prime = \frac{U}{2} \sigma_{j_0}^z,
\label{eq:Hprime_Pauli}
\end{equation}
leaving out the constant term. The matrices $\sigma^a$ can be transformed into bond operators by the Kramers-Wannier transformation \cite{Kramers_Statistics_1941, Essler_Duality_2009} 
\begin{align}
\begin{split}
\tau_{j + \frac{1}{2}}^z &= \sigma_j^z \sigma_{j + 1}^z, \\
\tau_{j + \frac{1}{2}}^y &= \prod_{i = 1}^j \sigma_i^x.
\end{split}
\label{eq:bond_operators}
\end{align}
The indexing of the bonds is illustrated in Fig.~\ref{fig:bond_indices}. When $\sigma^y = -i \sigma^z \sigma^x$ is also transformed, the Hamiltonian (\ref{eq:H0_Pauli}) becomes
\begin{equation}
H_b = \frac{J_{xy}}{4} \sum_{j = 2}^{L - 2} \left(1 - \tau_{j + \frac{1}{2}}^z \right) \tau_{j - \frac{1}{2}}^y \tau_{j + \frac{3}{2}}^y + \frac{J_{xy} \Delta}{4} \sum_{j = 1}^{L - 1} \tau_{j + \frac{1}{2}}^z
\label{eq:Hamiltonian_tau}
\end{equation}
and $\sigma_{j_0}^z$ in term (\ref{eq:Hprime_Pauli}) becomes
\begin{equation}
\sigma_{j_0}^z = \sigma_{1}^z \prod_{j = 1}^{j_0-1} \tau_{j + \frac{1}{2}}^z.
\label{eq:sigma_z}
\end{equation}
for $j_0 \geq 2$. We consider the case where the system is initially in the commensurate Mott insulator phase and $\Delta \gg 1$. When the number of lattice sites is odd and $\sum_j \langle S_j^z \rangle = \frac{1}{2}$, the spins at the edge sites are up and one can fix the boundary condition $\langle \sigma_1^z \rangle = 1$.

If neighboring spins point in opposite directions, the bond operator $\tau^z$ gives the value $-1$, and for neighboring spins in the same direction, $+1$. The value $+1$ therefore corresponds to the existence of a domain wall, which can be a soliton (two neighboring spins up) or an antisoliton (two neighboring spins down). We will use here the term domain wall (DW) for both since they have the same bond representation. For the configuration of Fig.~\ref{fig:bond_indices}, the value of $\langle \sigma_{j_0}^z \rangle$ is determined by the parity of $j_0$. When $j_0$ is odd, $\langle \sigma_{j_0}^z \rangle = 1$, and when $j_0$ is even, $\langle \sigma_{j_0}^z \rangle = -1$. We focus on the case where $j_0$ is odd and there is initially a spin up at $j_0$.

\begin{figure}
\begin{center}
	\includegraphics[width=\linewidth]{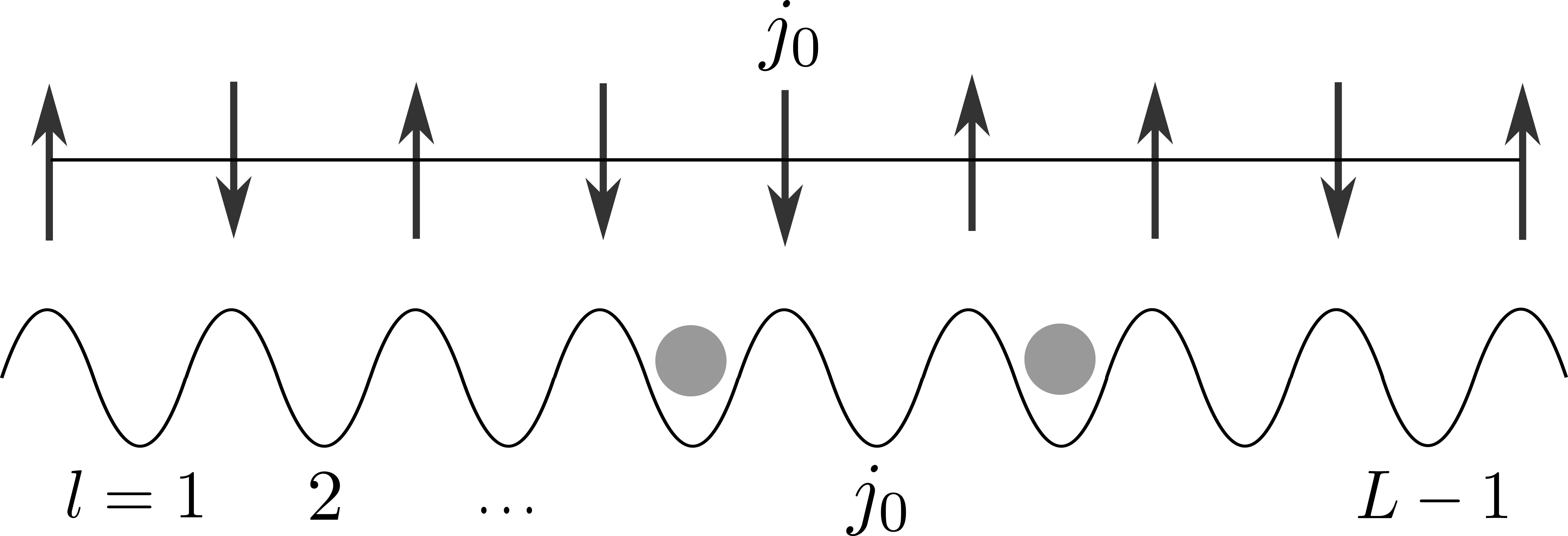}
	\caption{Instead of lattice sites, the system can be represented as bonds. The bonds are either empty or occupied by a spinless fermion, corresponding to a domain wall in the spin chain.}
	\label{fig:dual_lattice}
\end{center}	
\end{figure}

Changing into the bond indices $l = j - \frac{1}{2}$, the operators $\tau_l^z$ and $\tau_l^y = i \left( \tau_l^- - \tau_l^+ \right)$ can be transformed into spinless fermion operators using
\begin{align*}
\tau_l^z &= 2 c_l^\dagger c_l - 1, \\
\tau_l^+ &= c_l^\dagger e^{-i \pi \sum_{m < l} c_m^\dagger c_m}.
\end{align*}
In terms of the fermion operators, one can write \cite{Essler_Duality_2009}
\begin{align}
\begin{split}
H_b &= \frac{J_{xy}}{2} \sum_{l = 2}^{L - 2} [ \left( 1 - c_l^\dagger c_l \right) c_{l - 1}^\dagger c_{l + 1} \\
&- \left( 1 - c_l^\dagger c_l \right) c_{l - 1}^\dagger c_{l + 1}^\dagger + \text{H.c.} ] + \frac{J_{xy} \Delta}{2} \sum_{l = 1}^{L - 1} c_{l}^\dagger c_{l}
\end{split}
\label{eq:H0_fermions}
\end{align}
and
\begin{equation}
\sigma_{j_0}^z = \prod_{l = 1}^{j_0 - 1} \left( 2 c_{l}^\dagger c_{l} - 1\right).
\label{eq:Hprime_fermions}
\end{equation}
The creation of a fermion now corresponds to the creation of a DW in the original spin chain, as depicted in Fig.~\ref{fig:dual_lattice}. The DW representation can be applied to calculating both ground state quantities and time-dependent ones, as is done in Secs. \ref{sec:ground_state_density} and \ref{sec:time_evolution_density}.

\subsection{Density distribution in the ground state}
\label{sec:ground_state_density}

The density operator can be written in terms of the $S_j^z$ spin operator as $n_j^b = S_j^z + \frac{1}{2}$. One can directly use the expression (\ref{eq:Hprime_fermions}) for calculating the local magnetization $\langle S_j^z \rangle$ in a system with only few domain walls. The energy cost of creating a domain wall is large for $\Delta \gg 1$ and one can expect $\langle c_{l - 1}^\dagger c_{l + 1}^\dagger \rangle~\approx~0$. In Hamiltonian (\ref{eq:H0_fermions}), the first term on the second line can thus be neglected in this limit. When there is no creation or annihilation of domain walls, the last term of Eq. (\ref{eq:H0_fermions}) is a constant shift in energy and can also be left out. 

When the density of domain walls is very low, $\langle c_l^\dagger c_l \rangle~\approx~0$, one can approximate 
\begin{equation}
\left(1 - c_l^\dagger c_l \right) \approx 1
\end{equation}
on the first line of Eq. (\ref{eq:H0_fermions}). This approximation removes the interaction and leads to the free fermion Hamiltonian
\begin{equation}
H_b \approx \frac{J_{xy}}{2} \sum_{l = 2}^{L - 2} \left( c_{l - 1}^\dagger c_{l + 1} + \text{H.c.} \right).
\label{eq:noninteracting_approximation}
\end{equation}
This Hamiltonian is an excellent approximation of (\ref{eq:H0_fermions}) as long as the domain walls are far from each other. However, using Hamiltonian (\ref{eq:noninteracting_approximation}) directly to compute the ground state density distribution would be inadequate. It separates into two decoupled Hamiltonians for the odd and even bonds, $H_b = H_{\text{odd}} + H_{\text{even}}$.  The Hamiltonian $H_b$ therefore has degenerate pairs of eigenstates corresponding to $H_{\text{odd}}$ and $H_{\text{even}}$, with nearly identical wavefunctions. The wavefunctions corresponding to two eigenstates with the same energy have a similar envelope, but the ones which correspond to the eigenstates of $H_{\text{odd}}$ are only nonzero on the odd bonds, and respectively for $H_{\text{even}}$. In this description, the domain walls in the two degenerate states are discernible, and fermionic statistics do not exclude two domain walls in the same region of space. In Hamiltonian (\ref{eq:H0_fermions}), the interaction on the first line prevents the DWs from crossing and ensures the correct fermionic statistics.

When the number of domain walls is small, their wavelength is very large.
We can therefore take the continuum limit and ignore the microscopic lattice, since the envelopes of the wavefunctions vary at a much larger lengthscale. In this limit, we can consider the non-degenerate eigenstates of a particle in a box,
\begin{equation}
\varphi_k^l = \sqrt{\frac{2}{L}} \sin(k l),
\end{equation}
with the momenta $k = k_m = \frac{m \pi}{L}$, $m = 1, 2, \cdots, L$. We thus consider indistinguishable domain walls which can be located on all bonds. Using these states excludes properly two domain walls from the same region of space by the effect of fermionic statistics even without the interaction term.

The fermion operators have the momentum representation
\begin{equation}
c_l^\dagger = \sum_k \varphi_k^{l*} c_k^\dagger,
\label{eq:Fourier_transform}
\end{equation}
Substituting Eq. (\ref{eq:Fourier_transform}), the Hamiltonian of Eq. (\ref{eq:noninteracting_approximation}) can be written as
\begin{equation}
H_b = \sum_k \epsilon_k c_k^\dagger c_k,
\label{eq:H0_momentum}
\end{equation}
where $\epsilon_k = J_{xy} \cos(2 k)$. For this dispersion, there is a maximum of energy at $k = 0$. It is more convenient to have the minimum of energy at $k = 0$, and therefore we shift the momentum by $\frac{\pi}{2}$:
\begin{equation}
c_l^\dagger \rightarrow e^{- i \frac{\pi}{2} l} c_l^\dagger,
\end{equation}
which changes $H_b \rightarrow -H_b$. The basis functions shift as
\begin{equation}
\varphi_k^l = e^{- i \frac{\pi}{2} l} \sqrt{\frac{2}{L}} \sin(k l).
\end{equation}
The phase factor in $\varphi_k^l$ cancels in expectation values which contain $|\varphi_k^l|^2$. In Hamiltonian (\ref{eq:H0_momentum}), the dispersion relation becomes
\begin{equation}
\epsilon_k = -J_{xy} \cos(2 k).
\label{eq:dispersion}
\end{equation}

The ground state of the free Hamiltonian is the many-body state of $N$ noninteracting fermions,
\begin{equation}
\begin{split}
\ket{\Psi_{1, 2, \cdots, N}} = \sum_{l_1, \cdots, l_N} \varphi_{k_1, \cdots, k_N}^{l_1, \cdots, l_N} \ket{l_1} \otimes \ket{l_2} \otimes \cdots \ket{l_N}.
\end{split}
\end{equation}
Here, $l_{\alpha}$ denotes the coordinate of fermion $\alpha$. In the direct product, $\ket{l}$ denotes a single-particle state where site $l$ is occupied and other sites are empty,
\begin{equation}
\ket{l} = \ket{0, \cdots, 0, 1_l, 0, \cdots, 0}.
\end{equation}
The coefficient $\varphi_{k_1, \cdots, k_N}^{l_1, \cdots, l_N}$ is given by the Slater determinant
\begin{equation}
\varphi_{k_1, \cdots, k_N}^{l_1, \cdots, l_N} = \frac{1}{\sqrt{N!}}
{\begin{vmatrix}
\varphi_{k_1}^{l_1}	&\varphi_{k_2}^{l_1}	&\cdots		&\varphi_{k_N}^{l_1} \\
\varphi_{k_1}^{l_2}	&\varphi_{k_2}^{l_2}	&\cdots		&\varphi_{k_N}^{l_2}	\\
\vdots				&\ddots				&			&\vdots			\\
\varphi_{k_1}^{l_N}	&\cdots				&			&\varphi_{k_N}^{l_N}
\end{vmatrix}}.
\end{equation}
Equation (\ref{eq:Hprime_fermions}) can be used for calculating the expectation value of magnetization $\langle S^z_j \rangle = \frac{1}{2} \langle \sigma_j^z \rangle$:
\begin{align}
\begin{split}
&\bra{\Psi_{1, \cdots, N}} S^z_j \ket{\Psi_{1, \cdots, N}} \\
&= \frac{1}{2} \bra{\Psi_{1, \cdots, N}}  \prod_{d = 1}^{j - 1} \left( 2 c_{d}^\dagger c_{d} - 1\right)\ket{\Psi_{1, \cdots, N}} \\
&= \frac{1}{2} \sum_{l_1, \cdots, l_N} |\varphi_{k_1, \cdots, k_N}^{l_1, \cdots, l_N}|^2 \prod_{d = 1}^{j - 1} \left[ 2 \sum_{\alpha = 1}^N \delta_{d, l_{\alpha}} - 1 \right],
\end{split}
\label{eq:magnetization_DW}
\end{align}
which gives the expectation value of density as 
\begin{equation}
\langle n_j^b \rangle = \langle S_j^z \rangle + \frac{1}{2}.
\label{eq:density}
\end{equation}

\begin{figure}
\begin{center}
	\includegraphics[width=\linewidth]{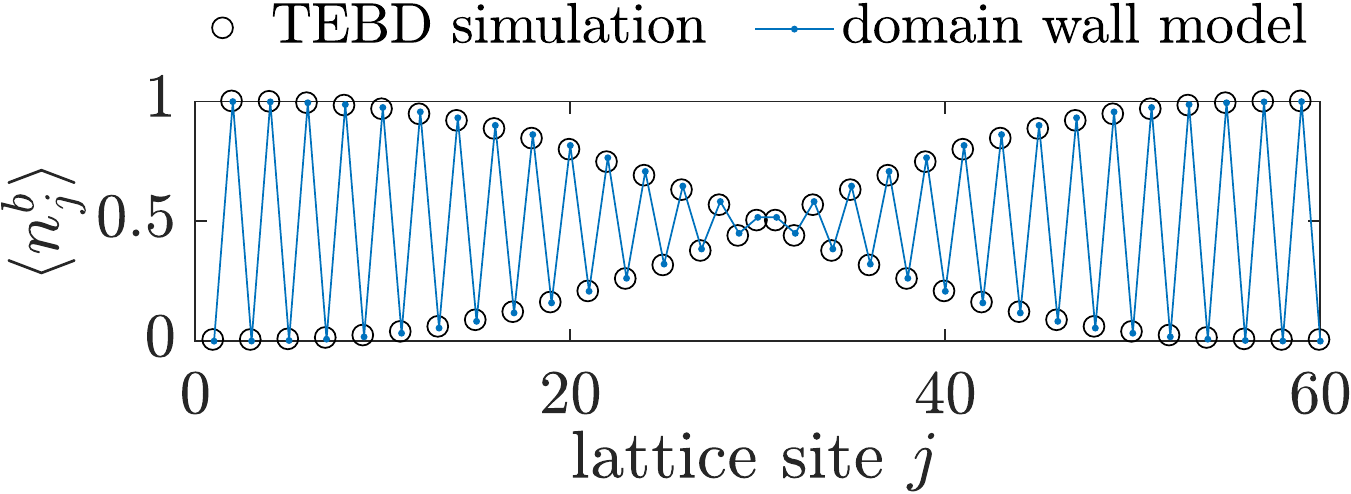}
	\caption{The density distribution has one domain wall when $L = 60$ and the number of fermions is $N_b = \frac{L}{2} + 1$. The result of the DW model for $N = 1$ agrees very well with the many-body TEBD solution.}
	\label{fig:one_soliton_comparison}
\end{center}	
\end{figure}

\begin{figure}
\begin{center}
	\includegraphics[width=\linewidth]{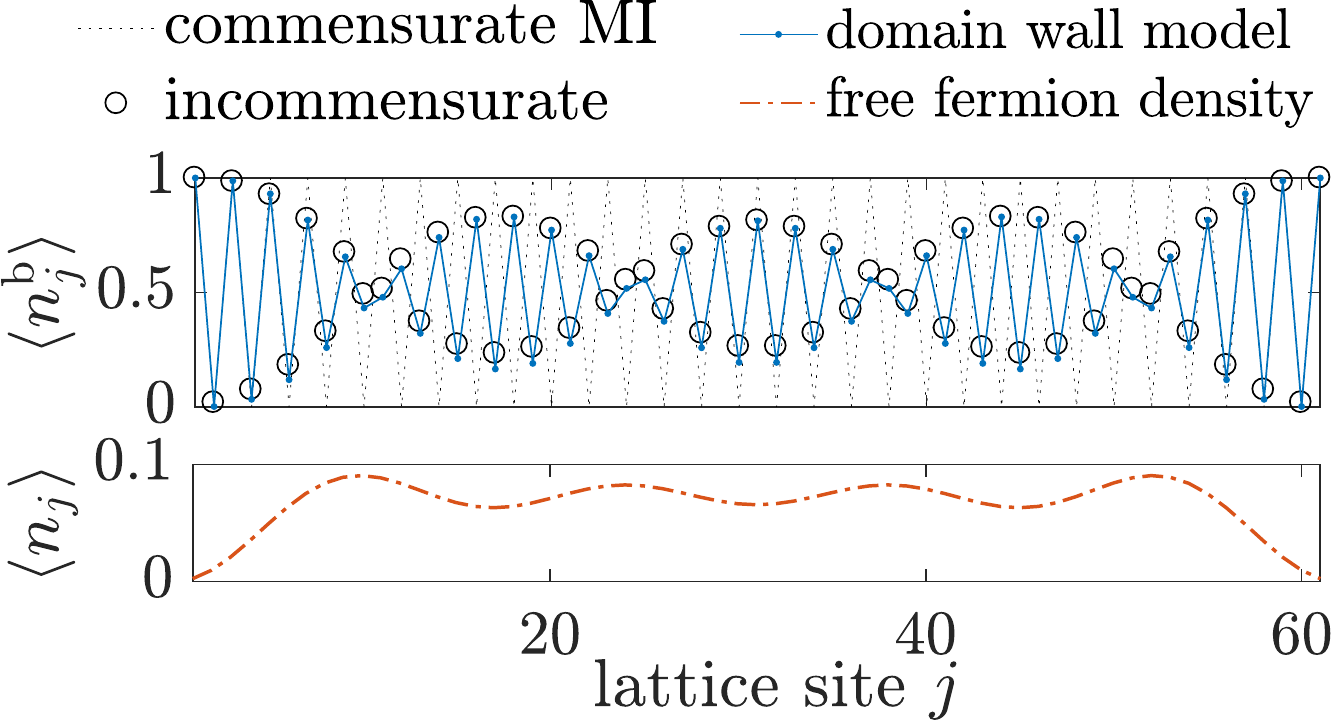}
	\caption{Upper panel: The ground state density distribution in the MI and incommensurate phases. The nearest-neighbor interaction is $V = 50 J$ and the size of the lattice is $L = 61$. The number of fermions is $N_b = 31$ in the MI case and $N_b = 33$ in the incommensurate case, which gives $N_{\text{DW}} = 4$. Lower panel: The density distribution of free fermions $\langle n_j \rangle$ as in Eq.~(\ref{eq:free_fermion_density}) with $N = 4$.}
	\label{fig:ground_state_supp}
\end{center}	
\end{figure}

The number of fermions $N$ is equal to the number of domain walls $N = N_{\text{DW}}$. In the ground state, the number of DWs is given by the number of bath fermions $N_b$, 
\begin{equation}
N_{\text{DW}} = 2 \left( N_b - \frac{L}{2} \right) - 1.
\end{equation}
In Fig. \ref{fig:ground_state_supp} with $\left( N_b - \frac{L}{2} \right) = 2.5$, there are four DWs. The most probable locations of the DWs are seen as two neighboring maxima or minima in the density distribution of the upper panel. The result of Eq.~(\ref{eq:density}) and the many-body TEBD solution are shown in Fig. \ref{fig:one_soliton_comparison}
for $N = 1$ and $L = 60$. There is a very good agreement between the two solutions. When $N > 1$, the accuracy of the approximation (\ref{eq:noninteracting_approximation}) should reduce since there can be interactions between the DWs. For $N = 4$ and $L = 61$, we find however a good agreement between Eq. (\ref{eq:magnetization_DW}) and the numerical TEBD result, as shown in Fig.~\ref{fig:ground_state}(b) of the main text and in the upper panel of Fig.~\ref{fig:ground_state_supp}.

One can see the correspondence between solitons and free fermions also by comparing the many-body density distribution in the upper panel of Fig.~\ref{fig:ground_state_supp} directly to the density  distribution of free fermions 
\begin{equation}
\bra{FS} n_j \ket{FS} = \sum_k |\varphi_k^j|^2
\label{eq:free_fermion_density}
\end{equation}
shown in the lower panel of Fig. \ref{fig:ground_state_supp}. The sum is over $k =  \frac{\pi}{L + 1}, \cdots, k_F$, where $k_F = \frac{N \pi}{L + 1}$, and $\ket{FS} = \prod_{k \leq k_F} c_k^{\dagger} \ket{0}$. The positions of neighboring minima or maxima in the many-body density distribution coincide with the maxima of the free fermion density, which has Friedel oscillations due to the boundaries.

\subsection{Time evolution of the density distribution}
\label{sec:time_evolution_density}

In the Mott insulator phase, the time evolution of the density distribution after a local perturbation can be calculated in a similar way as the ground state density in Sec.~\ref{sec:ground_state_density}. We limit the discussion to the time evolution of one DW excitation. For incommensurate filling with more than one DW, one could not use the free-fermion description but would have to consider the interaction of the DWs.

As can be seen from Figs.~\ref{fig:bond_indices} and \ref{fig:dual_lattice}, if the up spin at site $j_0$ in the original lattice is exchanged with the down spin at site $j_0 + 1$, the DWs are created at bonds $j_0 - 1$ and $j_0 + 1$. If the up spin at site $j_0$ is exchanged with the down spin at site $j_0 - 1$, the DWs are created at bonds $j_0 - 2$ and $j_0$. When considering processes in the same energy sector, a DW created at bond $j_0 - 2$ or $j_0 - 1$ can only move on the left, $l < j_0$, and a DW created at bond $j_0$ or $j_0 + 1$ can only move on the right, $l \geq j_0$.
One can thus consider the time evolution of only half of the system. We consider an initial state where the DWs have already been created. The DW on the right side can be created at two neighboring bonds, $j_0$ or $j_0 + 1$. One can therefore write the initial state as a superposition
\begin{equation}
\begin{split}
\ket{\psi(0)} &= \frac{1}{\sqrt{2}} \left( \ket{j_0} + \ket{j_0 + 1} \right) \\
&= \frac{1}{\sqrt{2}} \sum_k \left[ \varphi_{k}^{j_0 *} + \varphi_{k}^{j_0 + 1 *} \right] \ket{k}.
\end{split}
\end{equation}
The calculation is presented here for the right side of the lattice. The left side is symmetric.

The time-dependent state is obtained by operating with the time evolution operator,
\begin{equation}
\begin{split}
\ket{\psi(t)} &= e^{-i H_b t} \ket{\psi(0)} \\
&= \frac{1}{\sqrt{2}} \sum_k \left[ \varphi_{k}^{j_0 *} + \varphi_{k}^{j_0 + 1 *} \right] e^{-i \epsilon_k t} \ket{k},
\end{split}
\end{equation}
where $H_b$ is given by eqs. (\ref{eq:H0_momentum}) and (\ref{eq:dispersion}). We operate on the time-evolved state with the operator $\sigma^z$ of Eq. (\ref{eq:Hprime_fermions})
\begin{equation}
\begin{split}
\sigma^z_j \ket{\psi(t)} = &\prod_{d = 1}^{j - 1} \left( 2 c_d^\dagger c_d - 1 \right) \frac{1}{\sqrt{2}} \sum_k \left[ \varphi_{k}^{j_0 *} + \varphi_{k}^{j_0 + 1 *} \right] \\
&\times e^{-i \epsilon_k t} \sum_{l = j_0}^L \varphi_{k}^l \ket{l} \\
= &\frac{1}{\sqrt{2}} \sum_{k, l} \left[ \varphi_{k}^{j_0 *} + \varphi_{k}^{j_0 + 1 *} \right] \varphi_{k}^l e^{-i \epsilon_k t} \\
&\times \prod_{d = 1}^{j - 1} \left( 2 \delta_{d, l} - 1 \right) \ket{l}.
\end{split}
\end{equation}
The local magnetization is now
\begin{equation}
\begin{split}
\bra{\psi(t)} S^z_j \ket{\psi(t)} 
= &\frac{1}{4} \sum_l \left| \sum_k \left[ \varphi_{k}^{j_0 *} + \varphi_{k}^{j_0 + 1 *} \right] \varphi_{k}^l e^{-i \epsilon_k t} \right|^2 \\
&\times \prod_{d = 1}^{j - 1} \left( 2 \delta_{d, l} - 1 \right).
\end{split}
\label{eq:DW_magnetization}
\end{equation}
Figure \ref{fig:symmetric_time_evolution} of the main text shows the density distribution given by Eq. (\ref{eq:DW_magnetization}) together with the many-body TEBD result. The initial state is slightly different in the TEBD simulation than in the DW model, which leads to some differences in the time evolution. The motion of the DW in the density is however quite accurately given by the simple model.

\section{Density correlation}
\label{app:density_correlation}

\begin{figure}
	\includegraphics[width=0.8\linewidth]{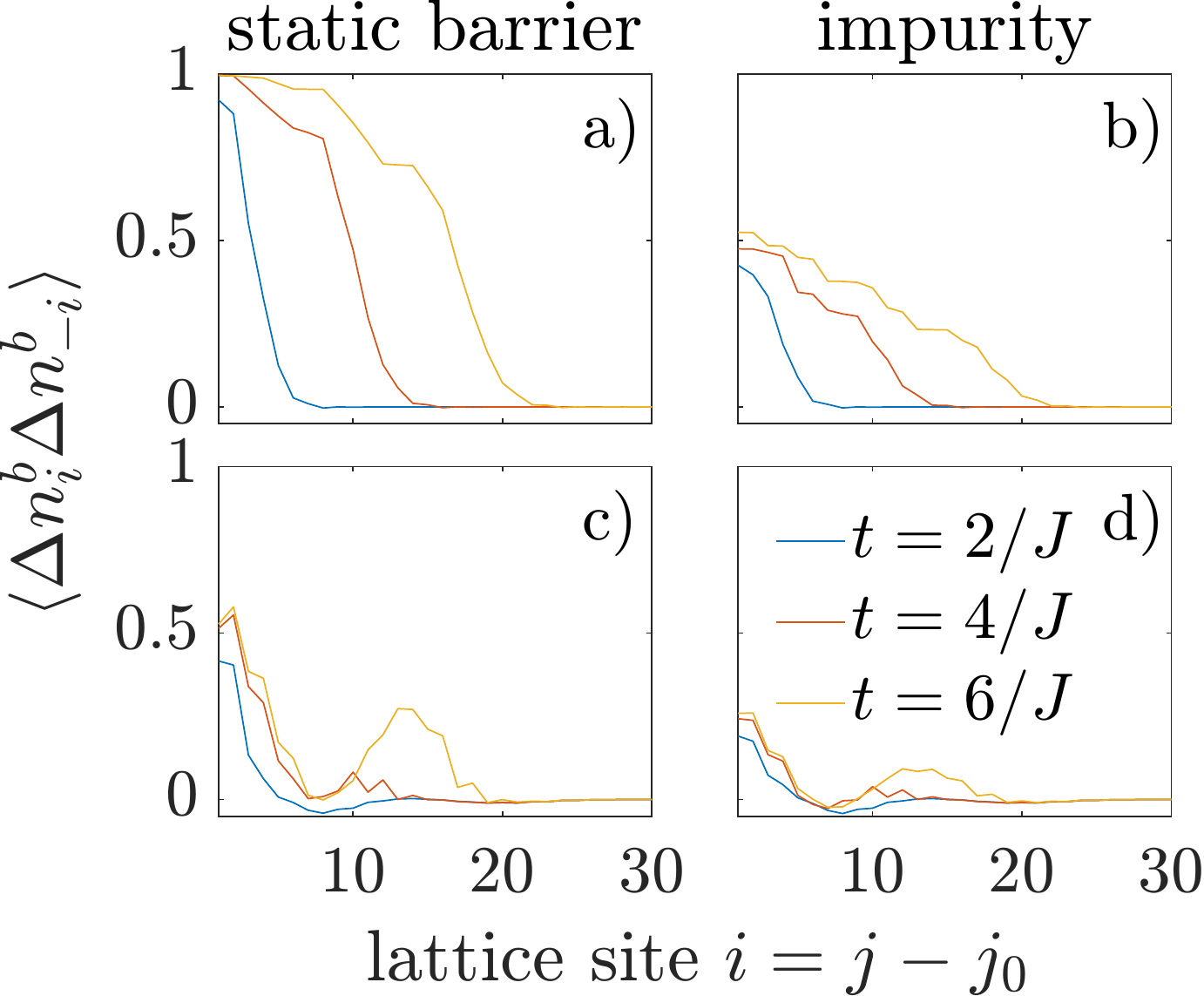}
    \caption{The correlator $\bra{\psi(t)} \Delta n^b_{i} \Delta n^b_{-i} \ket{\psi(t)}$ at different time steps in the case of a), c) a static potential barrier and b), d) an impurity. Panels a) and b) correspond to the MI and panels c) and d) to the incommensurate phase.}
    \label{fig:opposite_correlation}
\end{figure}

The complete inversion of density for the static barrier perturbation in Fig. \ref{fig:symmetric_time_evolution} a) of the main text suggests that the soliton and antisoliton excitation propagate in opposite directions with a probability close to 1. The soliton and antisoliton distributions of Figs. \ref{fig:commensurate_excitations} and \ref{fig:line_profiles} in the main text become zero at the center of the lattice, which is consistent with propagation to opposite directions. We can further investigate the symmetry in the motion of the excitations by computing the correlation
\begin{equation}
C_{i, -i}(t) = \bra{\psi(t)} \Delta n^b_i \Delta n^b_{-i} \ket{\psi(t)}
\label{eq:opposite_correlation}
\end{equation}
shown in Fig.~\ref{fig:opposite_correlation} (a).
Here, $i = j - j_0$ and
\begin{equation*}
\Delta n^b_i = n^b_i - \bra{\psi(0)} n^b_i \ket{\psi(0)}.
\end{equation*}
The value $C_{i, -i}(t) = 1$ means that the potential has produced an inversion of the magnetization at distance $i$ in a symmetric way, and the soliton and antisoliton excitations have propagated in opposite directions. A totally asymmetric time evolution where only one of the excitations propagates and the other one stays fixed would lead to $C_{i, -i}(t) = 0$. Values between 0 and 1 result from averaging over different possible time evolutions. These cases are detailed with schematic diagrams in Appendix C of \cite{Visuri2}.

The correlation in Fig.~\ref{fig:opposite_correlation} a) clearly indicates symmetric propagation. This in agreement with the DW model given in Sec.~\ref{sec:DW_model}, which indeed leads to the creation of a domain wall on both sides of the center site.
In the case of an impurity, the time evolution is more complicated. The impurity can move to the neighboring site and form a bound state with the antisoliton excitation, in which case only the soliton excitation propagates~\cite{Visuri2}. This results in a reduced amplitude of the magnetization oscillation with respect to the initial state, and the correlation in Fig. \ref{fig:opposite_correlation} (c) has a smaller maximum value than in Fig.~\ref{fig:opposite_correlation} (a). In both cases, however, one can see a light-cone propagation leading to zero correlation beyond a certain distance.

The correlation of Eq. (\ref{eq:opposite_correlation}) shown in Figs.~\ref{fig:opposite_correlation}(b) and \ref{fig:opposite_correlation}(d) has smaller values than in the commensurate case since the particles are initially partly delocalized. The maximum of the correlation is again smaller for the impurity than for the static barrier, since a similar bound state of the impurity and the possible antisoliton excitation can occur here as in the commensurate case. The correlation has an additional maximum between 10 and 20 sites from the center, corresponding to the density profile.

\section{Smaller interaction $V = 10 J$}
\label{app:smaller_interaction}

\begin{figure}
\begin{center}
	\includegraphics[width=\linewidth]{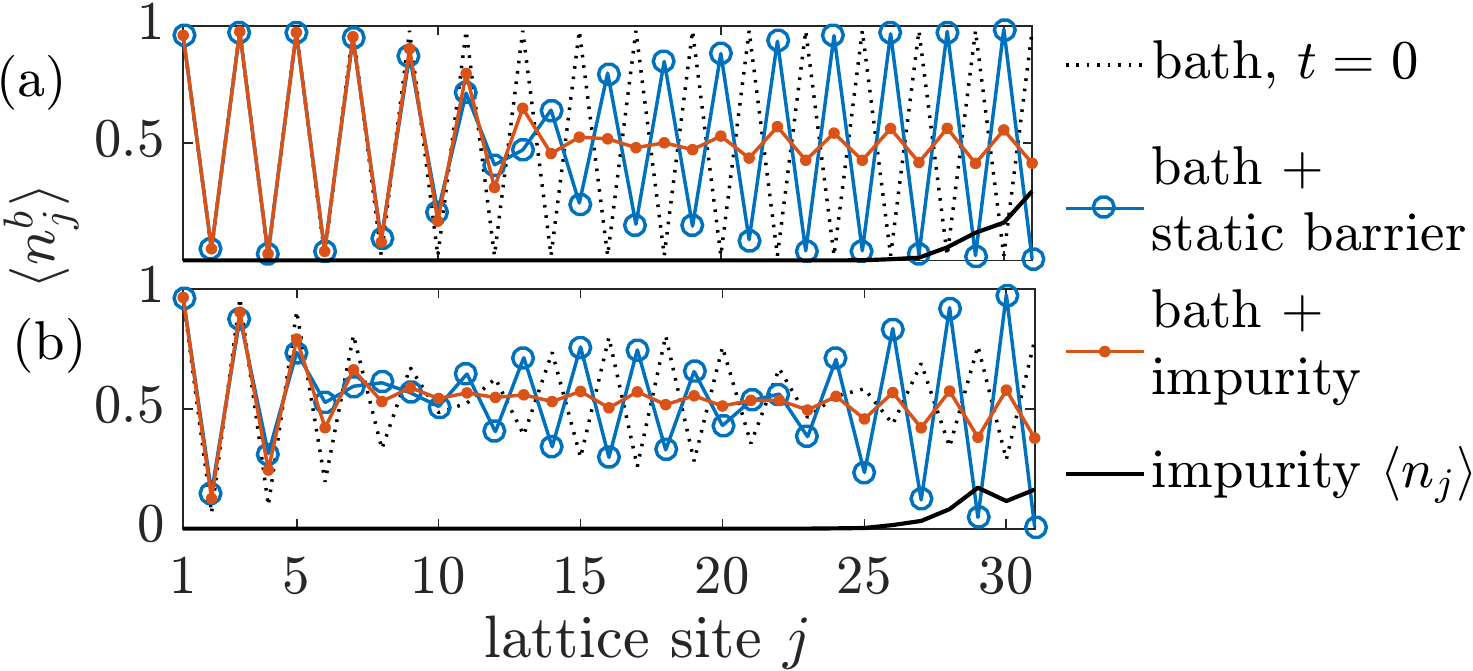}
	\caption{The density distribution of the bath fermions and the impurity in a) the commensurate and b) the incommensurate phase. The bath density is shown for both perturbations at times $t = 0$ and a) $t = 6/J$, b) $t = 8/J$. The nearest-neighbor and on-site interactions are $V = U = 10 J$.}
	\label{fig:density_V10}
\end{center}	
\end{figure}

For smaller interactions $U = V = 10 J$, we obtain essentially the same results as for $U = V = 50 J$. Due to a smaller nearest-neighbor repulsion, the particles are more delocalized in the initial state. This leads to a slightly reduced amplitude of the density oscillation in the ground state at $t = 0$ in Fig. \ref{fig:density_V10}.
Due to a larger effective tunneling $\frac{4 J^2}{U}$, the impurity moves further, which results in a broader maximum of the antisoliton distribution at the center of the lattice in Fig. \ref{fig:line_profiles} of the main text. Essentially the same features as for $U = V = 50 J$ can be seen in the time evolution of the soliton and antisoliton distributions in Fig.~\ref{fig:soliton_antisoliton_V10}:
\begin{figure}
\begin{center}
	\includegraphics[width=\linewidth]{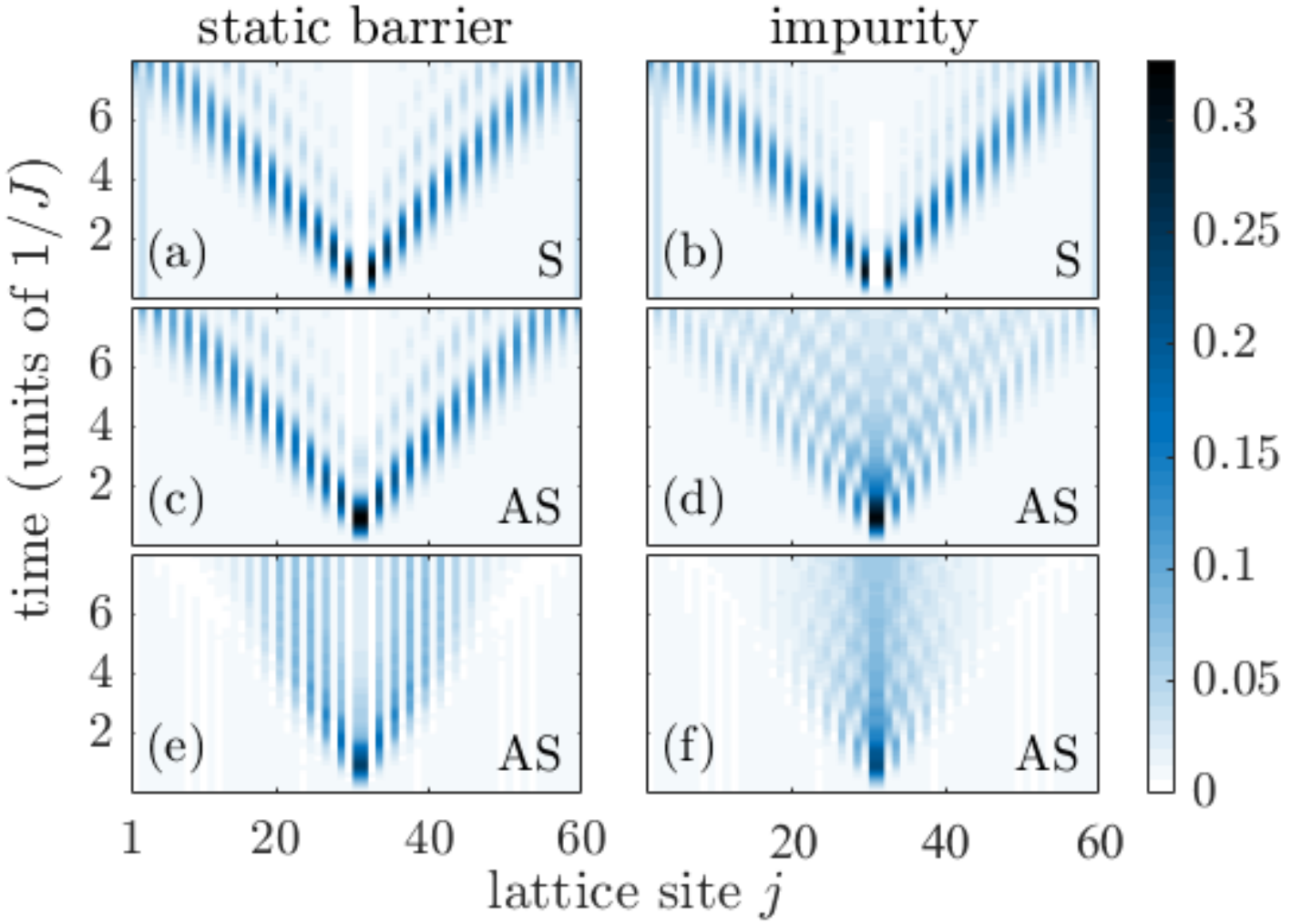}
	\caption{Panels a) and b) show the antisoliton distribution $\langle n_j^{\text{h}} n_{j + 1}^{\text{h}} \rangle$ as a function of position and time for the MI initial state when the system is perturbed with a static barrier or an impurity, respectively. Panels c) and d) show the corresponding soliton distributions $\langle n_j^{\text{b}} n_{j + 1}^{\text{b}} \rangle$. Panels e) and f) show the antisoliton distribution for incommensurate filling.}
	\label{fig:soliton_antisoliton_V10}
\end{center}	
\end{figure}
The excitations propagate away from the center when a static barrier is created, whereas for the impurity, the soliton propagates while the antisoliton distribution has a maximum at the center. For the incommensurate filling shown in panels e) and f), the antisoliton excitations clearly propagate a shorter distance than in the MI.

\begin{figure}
\begin{center}
	\includegraphics[width=0.85\linewidth]{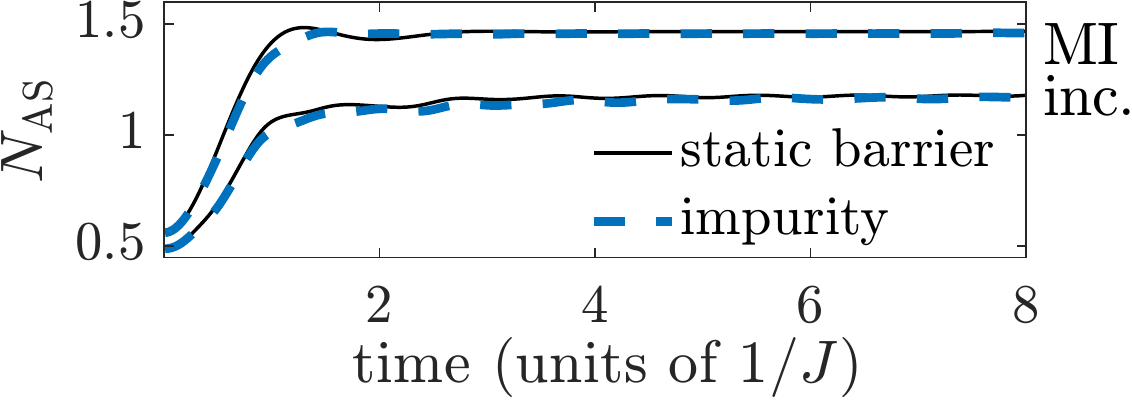}
	\caption{The number of antisolitons in the MI and incomensurate states for $U = V = 10 J$. The initial value is larger than for $U = V = 50 J$ since the energy cost of neighboring occupied sites in the ground state is smaller.}
	\label{fig:soliton_number_V10}
\end{center}	
\end{figure}

As seen in Fig. \ref{fig:soliton_antisoliton_V10}, there is a small density of antisolitons in the ground state. This is seen in the total number of antisolitons in Fig. \ref{fig:soliton_number_V10}, which is larger than for $U = V = 50 J$. The number of antisolitons at $t = 0$ and the increase during the simulation time are shown in Table \ref{table:antisoliton_numbers}. For $U = V = 50 J$, $N_{\text{AS}}$ is initially close to zero and the increase is approximately equal to the initial occupation of site $j_0$ in both the MI and incommensurate phases. The probability of not creating excitations is thus approximately equal to the probability of site $j_0$ being empty. For $U = V = 10 J$, the increase in the number of antisolitons is slightly smaller than the initial occupation of site $j_0$. When the particles are more delocalized, it is possible that both the sites $j_0$ and a site next to it are occupied. In this case, an additional soliton-antisoliton pair would not be created.
\begin{table}
\caption{The number of antisolitons in the initial state, the difference $\Delta N_{\text{AS}} = N_{\text{AS}}(t = 8/J) - N_{\text{AS}}(0)$, and the initial occupation probability of site $j_0$. The numbers are shown for the MI and incommensurate states and interactions $U = V = 10 J$ and $U = V = 50 J$. The number at time $t = 8/J$ corresponds to the time evolution with the local potential quench. For the impurity case, the numbers are the same within the accuracy shown here.}
\begin{center}
 \begin{tabular}{c c | c c c}
							&$U/J = V/J$				&$N_{\text{AS}}(0)$		&$\Delta N_{\text{AS}}$		&$\langle n_{j_0}^b (0)\rangle$ \\
\hline
\multirow{ 2}{*}{MI}		&50 	&0.02					&1							&1 \\
							&10 	&0.6					&0.9						&1 \\
							\hline
\multirow{ 2}{*}{incomm.}	&50 	&0.02					&0.8						&0.8 \\
							&10 	&0.5					&0.7						&0.8 \\
\end{tabular}	
\label{table:antisoliton_numbers}	
\end{center}
\end{table}

The presence of solitons can also be seen as DWs in the density distribution. The number of DWs $N_{\text{DW}}(t)$ in the density distribution is equal to the number of neighboring filled sites $N_{\text{S}}(t)$ for $V \rightarrow \infty$. For a finite $V$, more than $N_{\text{DW}}$ pairs of neighboring occupied sites can exist in the ground state due to delocalization, which leads to a smaller amplitude of the density oscillation. Correspondingly, the number of antisolitons in the ground state would be zero for $V \rightarrow \infty$ but can be nonzero for a finite $V$.  We have verified that $N_{\text{AS}}(t) \approx N_{\text{S}}(t)$ in the MI state, and $N_{\text{AS}}(t) \approx N_{\text{S}}(t) - N_{\text{DW}}(0)$ for incommensurate filling, with differences of order $10^{-3}$ for $U = V = 50 J$ and 0.08 for $U = V = 10 J$. We attribute the slightly larger soliton number to a small asymmetry in the system: a nondegenerate state for odd $L$ at half filling requires one filled site more than the number of empty sites.

\section{Harmonic trap}
\label{app:trap}

\begin{figure}
\begin{center}
	\includegraphics[width=\linewidth]{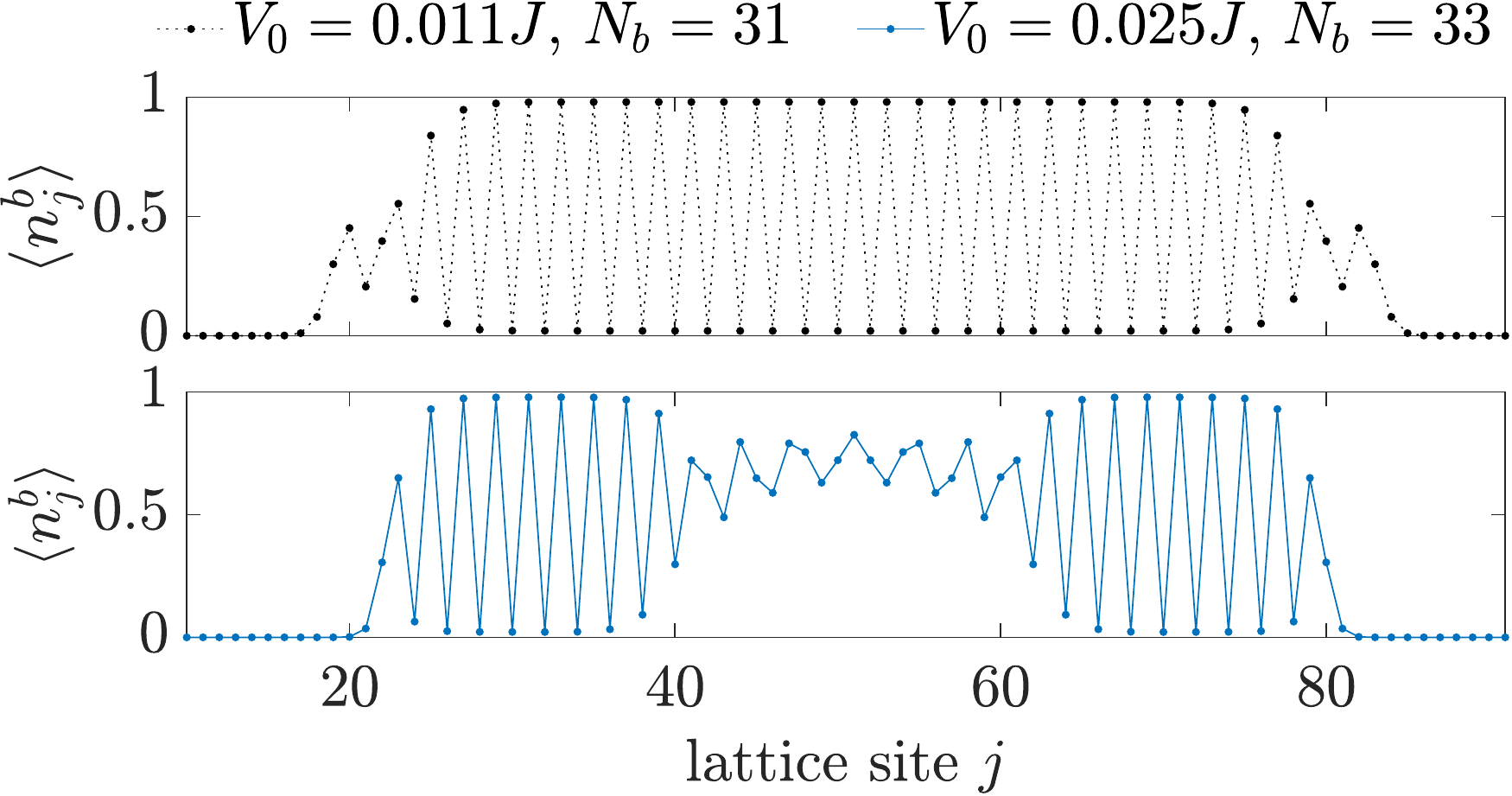}
	\caption{The density of particles in a harmonic trap, in the ground state of Hamiltonian (\ref{eq:Hamiltonian_total}). Upper: For a sufficiently shallow trap, the central region of the system is in the MI state. Lower: For a larger $V_0$ or $N_b$, the central region of higher average density is surrounded by Mott insulator regions. Here, $L = 80$, and the density distributions extend over approximately 60 sites.}
	\label{fig:density_large_lattice}
\end{center}	
\end{figure}

The ground state of the bath with a harmonic potential is described by the Hamiltonian 
\begin{equation}
H = H_{\text{b}} + H_{\text{trap}},
\label{eq:Hamiltonian_total}
\end{equation}
where
\begin{align}
H_{\text{trap}} = \sum_j V_{\text{trap}}(j) n^b_j
\label{eq:Hamiltonian_harmonic}
\end{align}
and $V_{\text{trap}}(j) = V_0 (j - j_0)^2$. For a sufficiently small $V_0$, i.e. when $V_0 (j - j_0)^2 \lesssim 2 V$ at $j = j_0 + N_b - 1$, the ground state in the central region of the trap is a Mott insulator, leading to the same dynamics as in a uniform system with open boundary conditions. The ground state density distribution is shown in Fig.~\ref{fig:density_large_lattice}.
For the ground state to be incommensurate, the particles must be confined by hard boundaries. The lower panel of Fig.~\ref{fig:density_large_lattice} shows the ground state density distribution for a stronger harmonic confinement, where the energy is minimized by a higher particle density in the center of the trap surrounded by Mott insulator regions. This distribution is qualitatively different from the incommensurate case of Fig.~\ref{fig:ground_state_supp}, and leads to different dynamics.

\begin{figure}
\begin{center}
	\includegraphics[width=0.9\linewidth]{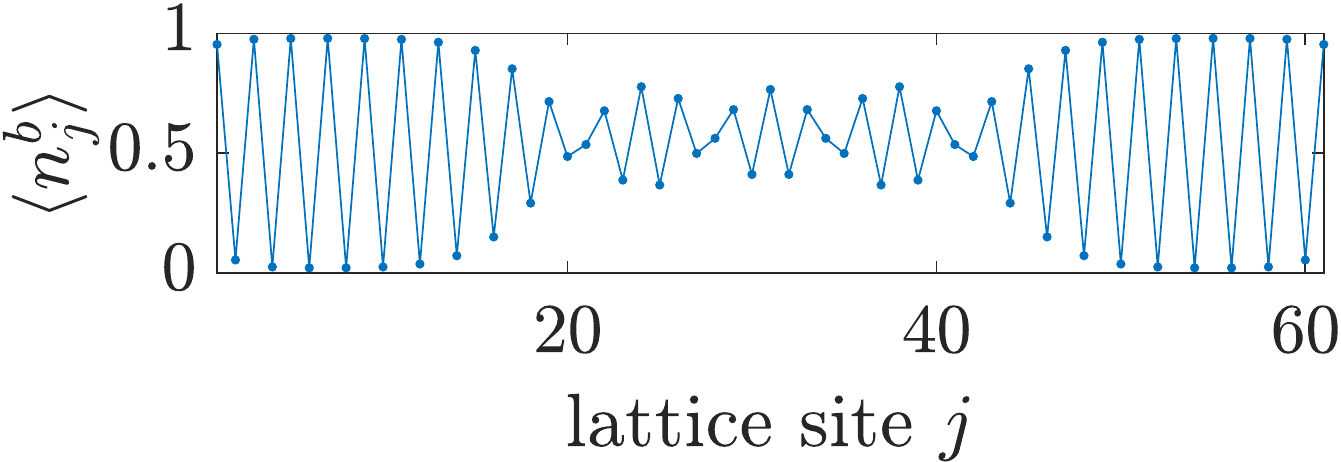}
	\caption{The density distribution in the ground state of Hamiltonian (\ref{eq:Hamiltonian_total}) calculated by TEBD. The strength of the harmonic potential $V_{\text{trap}}(j) = V_0 (j - j_0)^2$ is $V_0 = 0.005 J$. The filling is incommensurate with $N_b = 33$ and $L = 61$, which leads to four domain walls as in Fig.~\ref{fig:ground_state_supp}. The height of the potential at the edges of the lattice is $4.5 J$ and the nearest-neighbor repulsion is $V = 10 J$.}
	\label{fig:harmonic_density}
\end{center}	
\end{figure}

\begin{figure}
\begin{center}
	\includegraphics[width=\linewidth]{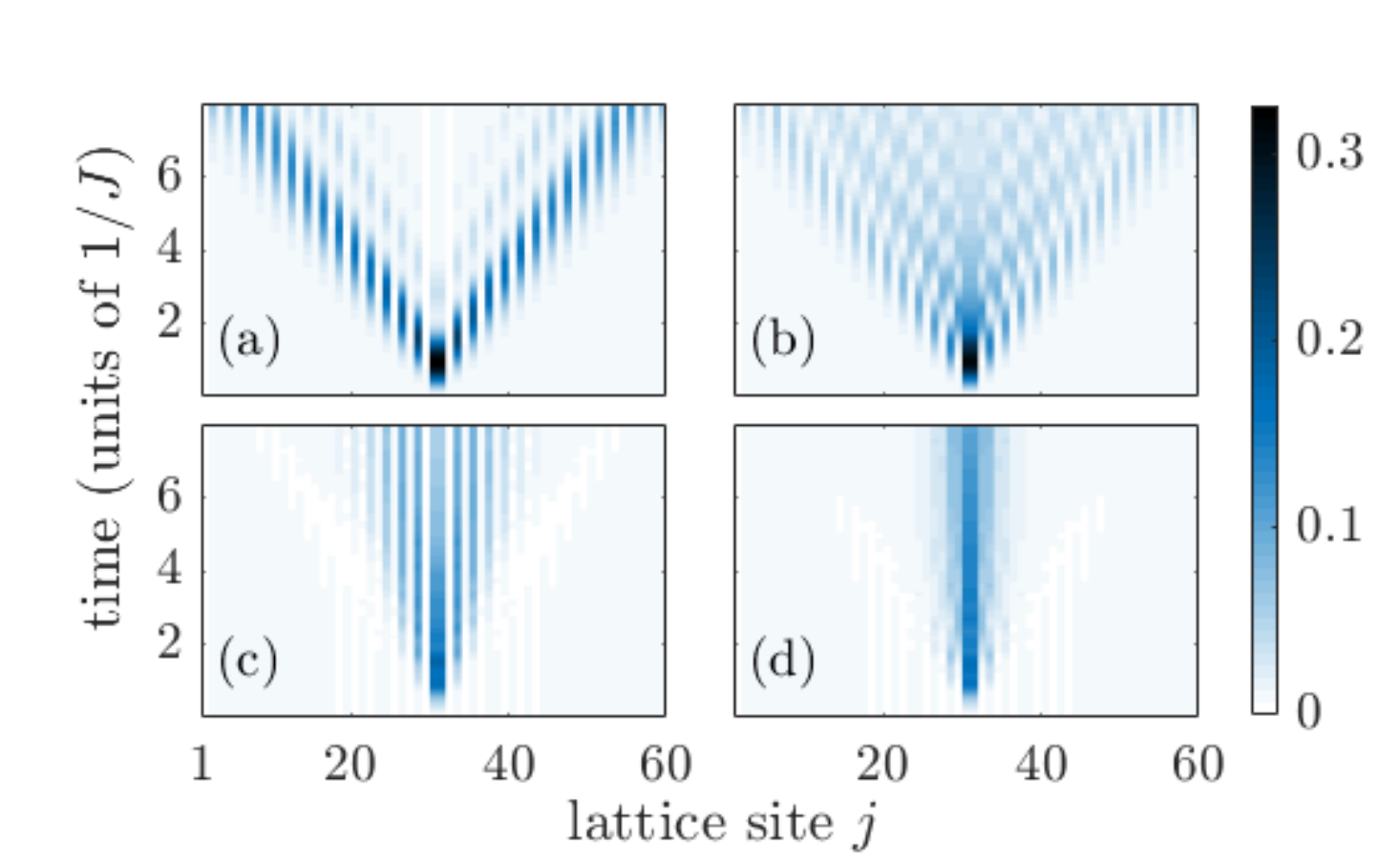}
	\caption{The antisoliton distributions as in Figs.~\ref{fig:soliton_antisoliton_V10}(c)--(f) in the case of a harmonic trapping potential with $V_0 = 0.005 J$. Here, $L = 61$, $N_b = 33$, and the nearest-neighbor repulsion is $V = 10 J$. At commensurate filling, the distribution is not significantly modified from the case of a uniform potential, whereas at incommensurate filling, the antisoliton is confined closer to the center of the lattice.}
	\label{fig:soliton_antisoliton_trap}
\end{center}	
\end{figure}

We find that when a shallow harmonic potential is superimposed on a system limited by hard boundaries, the particle density at incommensurate filling is similar to the one in a uniform box but the solitons are distributed closer to the center, as shown in Fig.~\ref{fig:harmonic_density}. 
At half filling, the ground state density distribution is not modified significantly by the harmonic potential when $V_{\text{trap}}(1) = V_{\text{trap}}(L) \ll 2 V$. 
In the time evolution, $H = H_{\text{b}} + H_{\text{trap}} + H^\prime$, where $H_{\text{trap}}$ is modified in the case of the impurity, 
\begin{equation}
H_{\text{trap}} = \sum_j V_{\text{trap}}(j) (n^b_j + n_j).
\end{equation} 
The time-dependent antisoliton distributions in Fig.~{\ref{fig:soliton_antisoliton_trap}} show that at incommensurate filling, the antisoliton is confined to a smaller region around $j_0$ than in Figs.~\ref{fig:soliton_antisoliton_V10}(e) and (f).

\bibliographystyle{unsrt}
\addcontentsline{toc}{section}{Bibliography}
\bibliography{bibfile}

\end{document}